\documentclass[a4paper, twocolumn, floatfix, 11pt, accepted=2024-08-19]{quantumarticle}
\pdfoutput=1
\usepackage[T1]{fontenc}
\usepackage[utf8]{inputenc}
\usepackage[english]{babel} 

\usepackage{amsmath}
\usepackage{physics}
\usepackage{verbatim}
\usepackage{graphicx}
\usepackage{xcolor}
\usepackage{setspace}
\usepackage{braket}
\usepackage{bbold}
\usepackage{color}
\newcommand\gfrac[2]{\genfrac{}{}{0pt}{}{#1}{#2}}
\usepackage[colorlinks,bookmarks=false,citecolor=blue,linkcolor=red,urlcolor=blue]{hyperref}
\usepackage[capitalise]{cleveref}
\Crefname{figure}{Fig.}{Figs.}
\Crefname{section}{Sec.}{Secs.}
\usepackage{nicefrac}
\usepackage{tipa}
\usepackage{amssymb}
\usepackage{wasysym}
\usepackage{bbold}
\usepackage[abs]{overpic}
\usepackage{cancel}
\usepackage{makecell}
\graphicspath{{Figures/}}

\makeatletter

 \definecolor{boxback}{HTML}{FFF8B5}
 
 \definecolor{applegreen}{rgb}{0, 0.5, 0.0}

\newcommand{\eqcontr}{\affiliation{These authors contributed equally to this work.}}

\newcommand{\padA}{\affiliation{Dipartimento di Fisica e Astronomia ``G. Galilei'', Università di Padova, I-35131 Padova, Italy.}}

\newcommand{\padB}{\affiliation{Padua Quantum Technologies Research Center, Università degli Studi di Padova}}

\newcommand{\padC}{\affiliation{Istituto Nazionale di Fisica Nucleare (INFN), Sezione di Padova, I-35131 Padova, Italy.}}

\newcommand{\bari}{\affiliation{Dipartimento di Fisica, Università di Bari, I-70126 Bari, Italy.}}

\newcommand{\ulm}{\affiliation{Institute for Complex Quantum Systems, Ulm University, Albert-Einstein-Allee 11, 89069 Ulm, Germany}}

\definecolor{smoothred}{HTML}{C5232F}
\definecolor{mygreen}{rgb}{0,0.5,0}
\definecolor{myblue}{rgb}{0,0,0.75}
\definecolor{mymagenta}{cmyk}{0,1,0,0.12}

\newcommand{\Id}{\mathbb{1}}

\newcommand{\avg}[1]{\left\langle#1\right\rangle}

\begin{document}

\title{Digital quantum simulation of lattice fermion theories with local encoding}

\author{Marco Ballarin}\eqcontr\padA \padC
\author{Giovanni Cataldi}\eqcontr\padA \padC
\author{Giuseppe Magnifico}\padA \padC \bari
\author{Daniel Jaschke} \padA \padC \ulm
\author{Marco Di Liberto} \padA \padC \padB
\author{Ilaria Siloi}\padA \padC \padB
\author{Simone Montangero}\padA \padC \padB
\author{Pietro Silvi}\padA \padC \padB

\begin{abstract}
We numerically analyze the feasibility of a platform-neutral, general strategy to perform quantum simulations of fermionic lattice field theories under open boundary conditions. 
The digital quantum simulator requires solely one- and two-qubit gates and is scalable since integrating each Hamiltonian term requires a finite (non-scaling) cost. The exact local fermion encoding we adopt relies on auxiliary $\mathbb{Z}_2$ lattice gauge fields by adding a pure gauge Hamiltonian term akin to the Toric Code. 
By numerically emulating the quantum simulator real-time dynamics, we observe a timescale separation for spin- and charge-excitations in a spin-$\frac{1}{2}$ Hubbard ladder in the $t-J$ model limit. 
\end{abstract}

\maketitle

\section{Introduction}
\begin{figure*}[t]
    \centering
    \includegraphics[width=\textwidth]{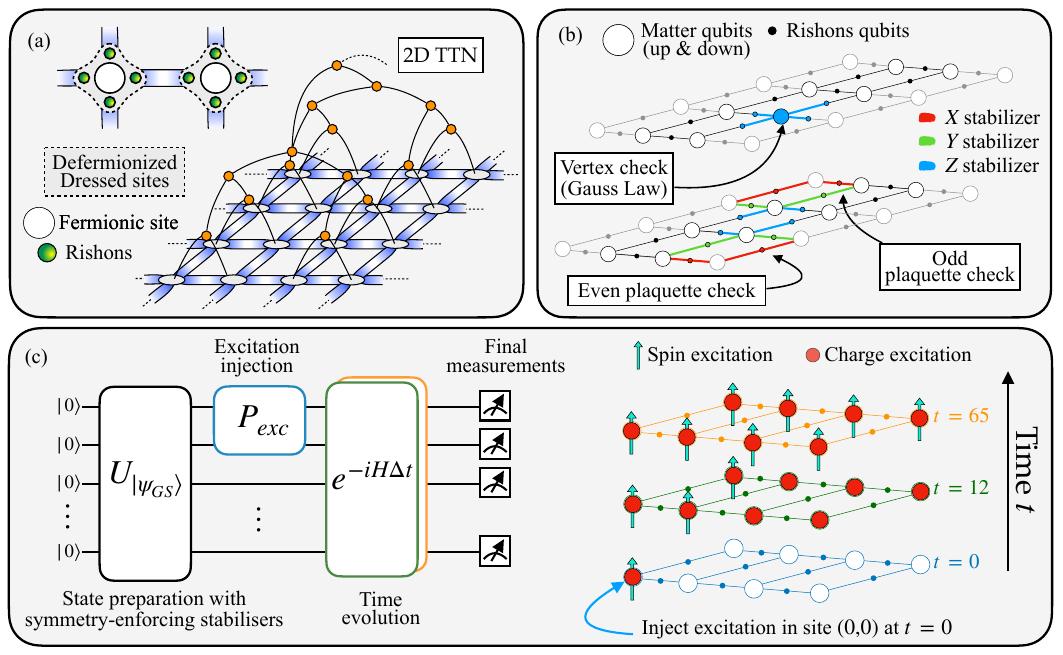}
    \caption{Simulation of the Hubbard model in 2D with tensor networks (equilibrium) and quantum circuits (out-of-equilibrium). \textbf{(a)} Schematic of the tree-tensor network (TTN) installed on the 2D lattice: the fermionic degrees of freedom are removed with the gauge defermionization and encoded in a dressed site (dashed closed line) comprising fermionic matter and rishons. With TTN, the ground state of the defermionized model is computed for up to $4\times4$ lattice. \textbf{(b)} To run the digital quantum simulation, dressed sites are decomposed into qubits and Hamiltonian terms in Pauli strings. The symmetries of the system are here enforced through the stabilizer formalism. \textbf{(c)} (left) Quantum circuit for simulating out-of-equilibrium dynamics: adiabatic preparation of the initial state (half-filling, repulsive $U$), injection of charge (spin) excitation, and time evolution with Hubbard Hamiltonian. (right) Spin-charge dynamics schematic: the charge excitation propagates faster than the spin one.}
    \label{fig_graph_abstract}
\end{figure*}

\noindent
The quest for quantum simulation of interacting fermionic models
\cite{fradkin2013, Auerbach, giamarchi2003} is considered a necessity to reach a novel understanding of collective phenomena both at low and high energies \cite{Lee_RevModPhys2006, Hartke2023a, FQHCollection, Hemery2023, Hofrichter2016}, but it is hindered by fundamental challenges \cite{Gattringer_IJMFA2016, Troyer_PRL2005, QS_roadmap21}. While analog quantum simulation, e.g.~with optical lattices, has advanced greatly in recent decades \cite{Esslinger_Rev2010, Bloch_NatPhys2012, Tarruel_Rev2018, Borhdt_Rev2021, Cheuk2015, Duarte2015, Edge2015, Greif2013, Haller2015, Hart2015, Hofstetter2002, Jordens2008, Messer2015, Murmann2015, Omran2015, Parsons2015, Schneider2008, Taie2012, Uehlinger2013}, it presents limitations in tailoring exotic interactions. And while fermionic digital quantum processors are still at an early stage of development \cite{TorstenDanielFermion1, TorstenDanielFermion2}, the well-established {\it conventional} digital quantum simulation platforms (e.g.~superconducting qubits, trapped ions, Rydberg arrays, quantum dots) \cite{arute2020observation, Barends_NatCommun2015, Sala_PRX2015, OMalley_PRX2016, Stanisic_NatCommun2022}
are built on distinguishable, spatially localized qubits (or qudits).
In this framework, {\it `Fermion Encoding'} is the analytical process of exactly converting a fermionic algebra (mutually-anticommuting operations) into a genuinely local algebra (mutually-commuting operations) of qudits.

Such encoding can not be carried out free of cost. Traditional strategies focused on encoding $N$ Dirac orbitals into $N$ qubits. 
By construction, these strategies can not preserve locality and, as a result, end up encoding the ubiquitous two-body interactions into cumbersome $W$-body interactions (with $W$ often addressed as \emph{Pauli weight}).
Alongside recent efforts, which were able to reduce $W$ from linear to logarithmic in $N$ \cite{JW1928,bravyi2022}, a separate sector of strategies arose: encodings attempting to preserve locality \cite{Fradkin80, Srednicki80, FermitospinFrank,chen2018}. 
They exhibit a flat Pauli weight ($W$ does not scale with $N$) at the price of requiring a number of qubits larger (but still linear scaling) than the number $N$ of fermion orbitals.
For non-gauge lattice fermion theories, Ref.~\cite{kitaev2006} showed that it is sufficient to add one qubit for each lattice bond to achieve local encoding (a generalization to lattice gauge fermionic theories was also recently introduced \cite{ErezA, ErezB}).
The extra qubits play the role of an effective, discrete lattice gauge field with pure gauge dynamics akin to the Toric Code Hamiltonian \cite{ToricCode}. 
The formal mapping is well understood \cite{chen_PRXQ2023}, but evidence of practical feasibility of the local encoding for numerical simulation or digital quantum simulation is still the subject of active research \cite{PardoPRR2023, MCB_Z2}.

In this work, we perform a twofold study on the feasibility of the local fermion encoding for classical and quantum simulation, as summarized in \cref{fig_graph_abstract}: on one hand side, we investigate its performance for ground-state numerical simulations based on tensor network ansatz states. 
On the other hand, we numerically test its applicability and scalability for digital quantum simulation on a generic platform (considering standard 1-qubit and 2-qubit programmable dynamical resources), by numerically emulating the noiseless quantum simulation processing in real-time.
Equipped with these methods, we investigate the properties of the spin-$\frac{1}{2}$ Hubbard model on a two-dimensional square lattice, both at equilibrium and out-of-equilibrium, via tensor networks. 
At zero temperature, we can access system sizes that allow us to identify the transition between liquid and insulating (crystalline spin-lattice) phases. 
In real-time dynamics for a ladder in the $t-J$ model limit, we observe spin and charge time-evolution displaying distinct time-scales, an effect that in 1D Hubbard chains is a precursor of the spin-charge separation phenomenon, see for example the recent experiments \cite{Hilker2017, arute2020observation, Vijayan_Science2020}. In contrast, in 2D, it only governs short time scales before the polaron picture and strongly-correlated effects set in \cite{Greiner2021, Bloch2021, Borhdt_Rev2021}.

The paper is organized as follows: in \cref{sec_defermion}, we revisit the gauge-based encoding to remove the fermionic matter locally, or gauge defermionization. 
We devote \cref{sec_TN} to implementing the encoding on a tree-tensor network (TTN) ansatz and show ground state simulation results of the 2D Hubbard model. 
Finally, in \cref{sec_fqm}, we detail our prescription for digital quantum simulation of the 2D Hubbard model, based on gauge defermionization, and numerically emulate the digital quantum simulation to observe spin-charge separation effects in two-dimensional lattice samples.

\section{Gauge Defermionization}
\label{sec_defermion}
In this section, we revisit and extend a known technique, based on lattice gauge theories, to eliminate fermionic matter from 2D lattice models \cite{ErezA, ErezB, Cataldi2024SimulatingSUYangMills}.
The strategy consists of developing an analytical mapping between an input fermion lattice Hamiltonian (here we consider it to be a pure non-gauge fermion theory) and a lattice gauge Hamiltonian, with $\mathbb{Z_2}$ gauge symmetry, equipped with appropriate gauge constraints. 
Furthermore, the resulting gauge theory can be manipulated so that the fermionic parity at each (dressed) site is protected, thus resulting in a theory where operator algebras at different sites always commute, i.e. a local spin theory.

Our formulation of the defermionization technique can be applied to any 2D fermion lattice Hamiltonian, regardless of the lattice system (it works on Honeycomb and Kagome lattices as well as irregular lattices), as long as few general constraints are satisfied:
\emph{(i)} the global parity of fermions is protected;
\emph{(ii)} Hamiltonian terms that flip the local fermion parity must be nearest-neighbor;
\emph{(iii)} the full system must be under open boundary conditions, i.e. every closed path on the lattice must be topologically shrinkable to a point.
These conditions are often satisfied in solid state Hamiltonians and cold atoms systems \cite{LiebBook}. 

Therefore, without any loss of generality, we consider a finite 2D lattice $\Lambda=(L_{x}, L_{y})$ whose sites and links are respectively identified by the couple ($\vb{j}, \vb*{\mu}$), where $\vb{j}=(j_{x},j_{y})$ is any 2D site, while $\vb*{\mu}$ is one of the two positive lattice unit vectors: $\vb*{\mu}_{x}=(1,0)$, $\vb*{\mu}_{y}=(0,1)$. Then, a general fermion lattice Hamiltonian reads:
\begin{equation}
\label{eq_originalhamiltonian}
\begin{split}
    H_{0} =& \sum_{\vb{j}\in \Lambda}\sum_{\vb*{\mu}}\sum_{f,f'}\qty(h_{f,f'} \, \psi^{\dagger}_{\vb{j},f} \psi_{\vb{j}+\vb*{\mu}, f'}+\text{H.c.}).
    \\
    &+\sum_{\vb{j}\in \Lambda}\sum_{\vb*{\mu}}\sum_{f, f'}\qty(\Delta_{f, f'} \, \psi_{\vb{j},f} \psi_{\vb{j}+\vb*{\mu}, f'} +
    \mbox{H.c.} \vphantom{ h_{f,f'} \psi^{\dagger}_{\vb{j},f}})\\
    &+ V\left( \{ n_{\vb{j},f} \}\right),
\end{split}
\end{equation}
where the fermion fields satisfy the usual Dirac anti-commutation relations
\begin{align}
    \qty{ \psi^{\dagger}_{\vb{j}, f}, \psi_{\vb{j'},f'}} &= \delta_{\vb{j},\vb{j'}} \delta_{f,f'}&
    \qty{ \psi_{\vb{j}, f}, \psi_{\vb{j'}, f'}} &= 0,
\end{align}
and can exhibit an internal degree of freedom, here labeled by the flavor index $f$. 

The interaction potential $V$ term, which solely is a function of the fermion densities $n_{\vb{j}, f} = \psi^{\dagger}_{\vb{j}, f} \psi_{\vb{j}, f}$, can in principle have any range and shape, and even include chemical potentials.
For instance, it can also account for local terms that do not conserve the spin, like $b_{f,f'}\psi^\dagger_{\vb{j},f}\psi_{\vb{j},f'}$, or local s-wave superconducting terms as $\Delta\psi^\dagger_{\vb{j},f}\psi^\dagger_{\vb{j},f'}$.

Conversely, the hopping $h_{f,f'}$ and the double creation/annihilation $\Delta_{f,f'}$ processes (the latter often seen in superconductors) are the terms that break local fermion parity: their single-site algebras do not commute between distant sites (i.e. they are not \emph{genuinely} local), and this is the source of all the numerical difficulty when simulating lattice fermions.

While in one spatial dimension, the Jordan-Wigner transformation provides an easy solution to this problem, tackling the mapping in higher dimensions requires sophisticated, often cumbersome techniques
\cite{FermitospinFrank,fermionicPEPS}. 
In this perspective, converting a fermion algebra into a genuinely local (spin-like) algebra using introducing a lattice gauge field is an elegant strategy, which is also practical from a numerical simulation standpoint, as we discuss later on.

\subsection{Step 1: Mapping into a $\mathbb{Z_2}$ gauge theory}
We now perform a set of exact algebraic manipulations to the Hamiltonian of \cref{eq_originalhamiltonian} until we reach a defermionized form. The very first step is to promote the total fermion parity operator at site $\vb{j}$ to a gauge transformation (for the matter sites $M$), namely \begin{equation}
    G_{\vb{j}}^{[M]} = \text{exp} \bigg(i \pi {\textstyle{\sum_f}}\psi^{\dagger}_{\vb{j}, f} \psi_{\vb{j}, f}\bigg).
\end{equation}
It is indeed a parity operator, or $\mathbb{Z}_2$ group, since 
\begin{equation}
    G_{\vb{j}}^{[M]} = G_{\vb{j}}^{[M] \dagger} = (G_{\vb{j}}^{[M]})^{-1}.
\end{equation}
Under $G_{\vb{j}}^{[M]}$, operators that preserve total fermion parity are left invariant, such as densities
$G_{\vb{j}}^{[M]} n_{\vb{j},f} G_{\vb{j}}^{[M]} = n_{\vb{j},f}$. Conversely, operators that flip the total fermion parity acquire a sign, that is $G_{\vb{j}}^{[M]} \psi_{\vb{j},f} G_{\vb{j}}^{[M]} = - \psi_{\vb{j},f}$.

At this stage, we promote the gauge transformation into a gauge symmetry. This task is performed by adding an auxiliary quantum lattice field, the gauge field, on the \emph{bonds} $(\vb{j},\vb{j}+\vb*{\mu})$
of the lattice. 
The local Hilbert space for a gauge site should correspond to the regular representation space for the $\mathbb{Z}_2$ group \cite{BurrelloZoharPRD}, so it should be a two-level system, or a qubit, equipped with the (genuinely local) algebra of Pauli matrices ${\vec{\sigma}}$. 

As $\mathbb{Z}_{2}$ is an Abelian gauge group, the left- $G_{\vb{j},\vb{j}+\vb*{\mu}}^{[L]}$ and right- $G_{\vb{j},\vb{j}+\vb*{\mu}}^{[R]}$ groups of transformations of the gauge field must commute and square to the identity, without loss of generality, we can set them both to the same operator: $G_{\vb{j},\vb{j}+\vb*{\mu}}^{[L]} = G_{\vb{j},\vb{j}+\vb*{\mu}}^{[R]} = \sigma^{z}_{\vb{j},\vb{j}+\vb*{\mu}}$.

Thanks to this auxiliary field, we equip the gauge-violating terms of the Hamiltonian with a parallel transporter operator. Namely, we replace
\begin{equation}
    \psi^{\dagger}_{\vb{j}, f} \psi_{\vb{j}+\vb*{\mu}, f'}\quad \longrightarrow \quad
    \psi^{\dagger}_{\vb{j}, f} U_{\vb{j},\vb{j}+\vb*{\mu}} \psi_{\vb{j}+\vb*{\mu}, f'},
\end{equation}
that transforms covariantly under the gauge field groups: 
\begin{equation}
    G_{\vb{j},\vb{j}+\vb*{\mu}}^{[L]}  U_{\vb{j},\vb{j}+\vb*{\mu}} G_{\vb{j},\vb{j}+\vb*{\mu}}^{[L]}{=}
G_{\vb{j},\vb{j}+\vb*{\mu}}^{[R]}  U_{\vb{j},\vb{j}+\vb*{\mu}} G_{\vb{j},\vb{j}+\vb*{\mu}}^{[R]}{=}- U_{\vb{j},\vb{j}+\vb*{\mu}}
\end{equation}
Again, without loss of generality, we set $U_{\vb{j},\vb{j}+\vb*{\mu}} = \sigma^{x}_{\vb{j},\vb{j}+\vb*{\mu}}$.

Overall, this procedure translates into modifying the fermion parity-flipping terms from \cref{eq_originalhamiltonian} according to
\begin{equation} 
\label{eq_modification}
  \begin{aligned}
   h_{f,f'} \, \psi^{\dagger}_{\vb{j}, f} \psi_{\vb{j}+\vb*{\mu}, f'}
   &\longrightarrow  &&
   h_{f,f'} \, \psi^{\dagger}_{\vb{j}, f} \sigma^{x}_{\vb{j},\vb{j}+\vb*{\mu}} \psi_{\vb{j}+\vb*{\mu}, f'}
\\
   \Delta_{f, f'} \, \psi_{\vb{j}, f} \psi_{\vb{j}+\vb*{\mu}, f'}
   &\longrightarrow  &&
   \Delta_{f,f'} \, \psi_{\vb{j}, f} \sigma^{x}_{\vb{j},\vb{j}+\vb*{\mu}} \psi_{\vb{j}+\vb*{\mu}, f'}\, .
  \end{aligned}
\end{equation}
With this modification, the Hamiltonian satisfies a gauge symmetry around every site,
precisely $G_{\vb{j}}^{[\text{total}]} = G_{\vb{j}}^{[M]} \prod_{\mu} G_{\vb{j},\vb{j}+\vb*{\mu}}^{[L/R]}$. Consequently, we have to choose a symmetry sector (for each of these symmetries) to play the role of physical space. 
As it is common for lattice gauge theories, we consider the quantum states \emph{invariant} under all the $G_{\vb{j}}^{[\text{total}]}$, i.e. which satisfy
\begin{equation} \label{eq_physpace}
 \left[\exp \left(i \pi {\textstyle{\sum_f}} \psi^{\dagger}_{\vb{j},f} \psi_{\vb{j}, f}\right)
 \prod_{\vb*{\mu}} \sigma^{z}_{\vb{j},\vb{j}+\vb*{\mu}}  \right]\ket{\Psi_{\text{phys}}}{=}
 \ket{\Psi_{\text{phys}}},
\end{equation}
in order to be the physical states $\ket{\Psi_{\text{phys}}}$. 
This equation plays the role of effective Gauss' Law of the resulting gauge theory, and the condition can be satisfied at every site \emph{only} if the total number of fermions in the system is even. To simulate an odd number of fermions, one or more virtual bonds, going out of the system, must be added to keep track of the appropriate parity gauge flux.

\subsection{Step 2: Restoring equivalence with Plaquette Operators}
It is then possible to check that, under the Gauss' Law of \cref{eq_physpace}, the gauge-theory modification \cref{eq_modification} does not change the actual dynamics given by the original Hamiltonian as in \cref{eq_originalhamiltonian}, \emph{but only} when the lattice is a non-cyclic graph (such as a 1D chain, or a dendrimer lattice). For each (product state) fermion configuration, the extra degrees of freedom introduced by the gauge fields are completely locked by Gauss' Law; thus all residual degeneracy is removed, and all the matrix elements of $H_0$ are unaltered.

Conversely, when lattice cycles are present, the modification changes the dynamics.
In fact, for each elementary closed cycle of sites, or \emph{plaquette} $\square\in \Lambda$, one Gauss' Law operator becomes linearly dependent on others, thus contributing to degeneracy with a 2-fold space. 
Moreover, when a fermion winds around a closed cycle, all of the gauge fields in its path undergo a $\sigma^{x}$ flip, thus the final state may differ from the original one.
If we want to restore the dynamics of the original model, we have to add more physical contents to the plaquette, either in terms of an additional \emph{pure gauge field} Hamiltonian or in terms of \emph{extra symmetries}.

Fortunately, a simple recipe to do either is inspired by Kitaev's Toric Code \cite{ToricCode}.
The idea is to add a plaquette Hamiltonian term with a $\sigma^{x}$ for each gauge link of the plaquette, that is 
\begin{equation}
\label{eq_plaquette_term}
\begin{split}
    H_p &=-\sum_{\square \in\Lambda}\prod_{\langle\vb{j},\vb{j'}\rangle \in \square}\sigma^x_{\vb{j},\vb{j'}}\\
    &=- \sum_{\square \in\Lambda}\qty(\begin{array}{ccc}
    \ulcorner& \sigma^{x} & \urcorner\\
    \sigma^{x} & & \sigma^{x} \\
    \llcorner&  \sigma^{x}   & \lrcorner\\
    \end{array}).
    \end{split}
\end{equation} 
Notably, each of these plaquette operators commutes with the lattice gauge Hamiltonian and can be equivalently cast as a plaquette symmetry. 
Moreover, plaquettes also commute with Gauss' laws, because a vertex and a plaquette always share either zero or two bonds, regardless of the lattice system, and $[\sigma^{x}_1 \sigma^{x}_2, \sigma^{z}_1 \sigma^{z}_2] = 0$. 

We can then map the original lattice fermion Hamiltonian, \cref{eq_originalhamiltonian}, to a dynamically-equivalent, $\mathbb{Z}_2$-invariant lattice gauge model, which reads
\begin{equation} \label{eq_torichamiltonian}
  \begin{split}
   H_1 &= \sum_{\vb{j},\vb*{\mu}}\sum_{f,f'}
    \left( h_{f,f'}\,\psi^{\dagger}_{\vb{j},f} \sigma^{x}_{\vb{j},\vb{j}+\vb*{\mu}} \psi_{\vb{j}+\vb*{\mu}, f'}
     \right.
    \\
    &\left. + \Delta_{f, f'} \, \psi_{\vb{j}, f} \sigma^{x}_{\vb{j},\vb{j}+\vb*{\mu}}  \psi_{\vb{j}+\vb*{\mu}, f'} +
    \mbox{H.c.} \vphantom{ h_{f, f'} \psi^{\dagger}_{\vb{j}, f}} \right)+ V\left( \{ n_{\vb{j}, f} \}\right),
  \end{split}
\end{equation}
with the constraints
\begin{equation} \label{eq_allgauges}
 \left\{ \begin{aligned}
    &\left( e^{i \pi \sum_f \psi^{\dagger}_{\vb{j}, f} \psi_{\vb{j}, f}}
 \prod_{\mu}^{\text{vertex}} \sigma^{z}_{\vb{j},\vb{j}+\vb*{\mu}} \right) \ket{\Psi_{\text{phys}}} = 
 \ket{\Psi_{\text{phys}}},
  \\
  &\left( \prod_{\avg{ \vb{j}, \vb{j'}}} ^{\text{plaquette}} \sigma^x_{\vb{j}, \vb{j'}} \right) \ket{\Psi_{\text{phys}}} = 
 \ket{\Psi_{\text{phys}}}.
  \end{aligned} \right.
\end{equation}
Indeed, the combined constraints in \cref{eq_allgauges} completely resolve the 2-fold degeneracy introduced by the plaquette, and each closed path of moving fermions returns to its original state with the correct amplitude and phase. Therefore, the whole mapping holds for the subspace of the Hilbert space satisfying \cref{eq_allgauges}. 
Any excitations from the ground state of \cref{eq_plaquette_term} break the encoding validity.
A rigorous proof of the equivalence between \cref{eq_originalhamiltonian} and Eqs.~\eqref{eq_torichamiltonian}-\eqref{eq_allgauges} was provided for a number-conserving theory on the square lattice in \cite{ErezA}; the generalization to other cases is fairly straightforward.

Notice that the degeneracy is not fully removed in periodic boundary conditions (PBC), where each winding dimension introduces an additional closed cycle. This case requires the additional constraint of a full $\sigma^{x}$ string along each winding dimension, but we will not treat this case here because the theory becomes non-local.

\subsection{Step 3: Fermionic Rishons}
So far, it seems that we have increased the complexity of the fermion lattice theory. In this last step, we will locally manipulate the gauge fields and achieve defermionization explicitly, ending with an algebra of genuinely local operators only. 

To do so, we split each gauge field, living on the $(\vb{j},\vb{j}+\vb*{\mu})$ lattice link, in a pair of spinless fermion modes (rishons) $c_{\vb{j},\vb*{\mu}}^{(\dagger)}$ and $c_{\vb{j}+\vb*{\mu},-\vb*{\mu}}^{(\dagger)}$, equipped with another symmetry. 
These new Dirac fermion operators belong to dressed sites $\vb{j}$ and $\vb{j}+\vb*{\mu}$ respectively and satisfy the usual anti-commutation relations between themselves
$\{ c^{\dagger}_{\vb{j}, \vb*{\mu}}, c_{\vb{j}', \vb*{\mu}'} \} = \delta_{\vb{j} j'} \delta_{\mu \mu'}$ and 
$\{ c_{\vb{j}, \vb*{\mu}}, c_{\vb{j}', \vb*{\mu}'} \} = 0$ and with the physical matter fermions $\qty{\psi_{\vb{j}},c_{\vb{j}',\vb*{\mu}}}=0$.
The combined 4-dimensional space of the two modes is then reduced back to the 2-dimensional space of a qubit by imposing that the total number of rishon fermions on a bond must be an even number. This constraint can again be cast as a link symmetry, requiring that 
\begin{equation}
    \exp[i \pi( c_{\vb{j},\vb*{\mu}}^{\dagger} c_{\vb{j},\vb*{\mu}}{+}c_{\vb{j}+\vb*{\mu},-\vb*{\mu}}^{\dagger} c_{\vb{j}+\vb*{\mu},-\vb*{\mu}})]{\ket{\Psi_{\text{phys}}}}{=}\ket{\Psi_{\text{phys}}}.
\end{equation}

We can now convert the Pauli algebra of the gauge fields into an operator algebra acting on the rishon pair: 
\begin{equation} \label{eq_sigmasplit}
  \begin{aligned}
\sigma^z_{\vb{j},\vb{j}+\vb*{\mu}} &\to 1 - 2 c_{\vb{j},\vb*{\mu}}^{\dagger} c_{\vb{j},\vb*{\mu}} = 1 - 2 c_{\vb{j}+\vb*{\mu},-\vb*{\mu}}^{\dagger} c_{\vb{j}+\vb*{\mu},-\vb*{\mu}} \\
\sigma^x_{\vb{j},\vb{j}+\vb*{\mu}} &\to i \gamma_{\vb{j},\vb*{\mu}} \gamma_{\vb{j}+\vb*{\mu},-\vb*{\mu}},
  \end{aligned}
\end{equation}
where now $\gamma_{\vb{j},\vb*{\mu}}{=}c_{\vb{j},\vb*{\mu}}{+} c^{\dagger}_{\vb{j},\vb*{\mu}}{=}\gamma_{\vb{j},\vb*{\mu}}^{\dagger}$ is a Majorana operator on the rishon fermion, which squares to the identity $\gamma_{\vb{j},\vb*{\mu}}^2{=}\Id$ but still anti-commutes with other fermions $\{\gamma_{\vb{j},\vb*{\mu}},\gamma_{\vb{j}',\vb*{\mu}'}\}{=}2 \delta_{\vb{j}, \vb{j'}} \delta_{\vb*{\mu},\vb*{\mu'}}$. 
These operators defined at \cref{eq_sigmasplit} respect the fermion parity symmetry on the bond, and on the even parity sector, they act exactly as Pauli matrices.

We can then plug this exact manipulation into the gauge theory Hamiltonian, resulting in
\begin{equation} 
\label{eq_defermionized}
  \begin{split}
   H&_2 = \sum_{\vb{j},\vb*{\mu}}\sum_{f,f'}
    \qty( i h_{f, f'} \, \psi^{\dagger}_{\vb{j}, f} \gamma_{\vb{j},\vb*{\mu}} \gamma_{\vb{j}+\vb*{\mu},-\vb*{\mu}} \psi_{\vb{j}+\vb*{\mu}, f'} +
    \text{H.c.})\\
    &+ \sum_{\vb{j},\vb*{\mu}}\sum_{f,f'}
    \qty(i\Delta_{f, f'} \, \psi_{\vb{j}, f} \gamma_{\vb{j},\vb*{\mu}} \gamma_{\vb{j}+\vb*{\mu},-\vb*{\mu}} \psi_{\vb{j}+\vb*{\mu}, f'} +
    \mbox{H.c.} \vphantom{ h_{f, f'} \psi^{\dagger}_{\vb{j}, f}} )\\
    &+ V\qty( \{ n_{\vb{j}, f} \}),
  \end{split}
\end{equation}
where we have to add the new link parity symmetry to the constraints, thus
\begin{equation} \label{eq_trigauges}
    \left\{\begin{aligned}
    &\left[ e^{i \pi \sum_f \psi^{\dagger}_{\vb{j}, f} \psi_{\vb{j}, f}}
 \prod_{\mu}^{\text{vertex}} e^{i \pi c^{\dagger}_{\vb{j}, \vb*{\mu}} c_{\vb{j}, \vb*{\mu}}}
\right]{\ket{\Psi_{\text{phys}}}}{=} 
 \ket{\Psi_{\text{phys}}}
  \\
  &\left( \prod_{\vb{j},\vb*{\mu}} ^{\text{plaquette}} i \gamma_{\vb{j},\vb*{\mu}} \gamma_{\vb{j}+\vb*{\mu},-\vb*{\mu}} \right) \ket{\Psi_{\text{phys}}} = 
 \ket{\Psi_{\text{phys}}}
 \\
 & e^{i \pi c_{\vb{j},\mu}^{\dagger} c_{\vb{j},\mu}} e^{i \pi c_{\vb{j}+\vb*{\mu},-\vb*{\mu}}^{\dagger} c_{\vb{j}+\vb*{\mu},-\vb*{\mu}} } \ket{\Psi_{\text{phys}}} =  \ket{\Psi_{\text{phys}}},
  \end{aligned}\right. 
\end{equation}
which must hold for every vertex, every plaquette, and every bond respectively.

We have finally reached the final form of our model:  now, every term of the Hamiltonian and every vertex, link, or plaquette symmetry \emph{protects total fermion parity} at each dressed site, counting together matter and rishon fermions (for plaquettes, remember that each dressed site contributes with two rishon modes in a closed lattice path).
Therefore, it is possible to think of each dressed site as a large spin, and both Hamiltonian and extra symmetries can be written in terms of \emph{genuinely local} operator algebras, which commute on different dressed sites. The theory has been effectively defermionized.

\subsection{The price of defermionization}
Defermionization of the lattice model using lattice gauge theory, i.e. going from
\cref{eq_originalhamiltonian} to Eqs.~\eqref{eq_defermionized}-\eqref{eq_trigauges} is an exact mapping. And while it provides the clear benefits of eliminating fermionic operator algebras, it does not come free of costs.

First of all, the transformation increases the local space dimension, only to shrink the local dimension again once the gauge symmetries are installed. 
If the model has $N$ sites, $f$ flavors and coordination number $v$,
we increase the dimension from $2^{fN}$ to $2^{(f+v)N}$.

Secondly, we impose a new interaction term in the form of the plaquettes. 
While for typical 2D lattices plaquettes are rather small, it is still an interaction involving three to six sites depending on the lattice system. If the original interaction $V$ was on-site, or nearest-neighbor, the defermionization effectively increased the interaction supports.
Moreover, there are no more components in the Hamiltonian that are quadratic in the Fermi operators (except chemical potentials within $V$). This means that Green's function perturbative approaches, which start from the free Fermi gas propagator of the quadratic theory, are no longer viable.

Finally, introducing an auxiliary field whose pure gauge dynamics is analogous to a Toric Code Hamiltonian has the drawback of increasing the entanglement. Even for product states of fermions, where fermions are locked in a specific integer filling configuration, and the original entanglement is zero (e.g.~for a Jordan-Wigner encoding), adding a Toric Code field on top of that raises the entanglement entropy to an exact area-law \cite{kitaev2009topological}, carrying one e-bit of entanglement per plaquette that is cut by the bi-partition. This observation has substantial implications for tensor network simulations of said models, as we will discuss in detail later on.

\section{Hubbard Model, Defermionized for tensor networks}
\label{sec_TN}
Despite its Hamiltonian's apparent simplicity, the Hubbard model has eluded physicists for decades \cite{Hubbard1963, HubbardReview2021}. While exact solutions are available for both one-dimensional \cite{Lieb_PRL1968} and infinite-dimensional cases \cite{Metzner_PRL1989,muller-hartmann_correlated_1989}, addressing finite-size systems in higher dimensions at arbitrary temperature has required the development of various computational techniques. 
Although Monte Carlo methods can handle substantially large systems, they suffer from the well-known sign problem when computing physically significant quantities within specific regimes. 
Comprehensive reviews detailing and comparing the accomplished computational outcomes for the 2D case can be found in the following references \cite{HubbardSim2021, LeBlanc_PRX2015}.

In this section, we join the effort by performing tensor network simulations of the Hubbard model at equilibrium at zero temperature. 
Here, we briefly review the model and manipulate its defermionized formulation to be ready for tensor network simulation. Subsequently, we will focus on a 2D \emph{square lattice} geometry, fermionic matter with 2 flavors (e.g. $\uparrow$ and $\downarrow$), and open boundary conditions.
On a rectangle lattice $\Lambda$ of $|\Lambda|=L_x \times L_y$ sites, the Hubbard Hamiltonian reads:
\begin{equation} 
\label{eq_Hubbard}
\begin{split}
   H_{\text{Hub}} =& 
   -t\sum_{\vb{j}\in\Lambda}\sum_{\vb*{\mu}}\sum_{f=\uparrow,\downarrow}
    \qty[\psi^{\dagger}_{\vb{j}, f} \psi_{\vb{j}+\vb*{\mu},f}^{} +\text{H.c.}]\\
    &+ U \sum_{\vb{j}\in\Lambda} \qty( n_{\vb{j},\uparrow} -\frac{1}{2})
    \qty( n_{\vb{j},\downarrow} -\frac{1}{2}),
  \end{split}
\end{equation}
where, apart from an energy rescaling, its ground state properties only depend on the dimensionless parameter $U/t$. 
The Hubbard model is regarded as the simplest theory of strongly correlated electrons, where band electrons interact via a two-body repulsive Coulomb interaction \cite{fradkin2013}. 
This model enables the description of a wide range of phenomena including metal-insulator transitions, superconductivity, and magnetism \cite{giamarchi2003}.  
The Hubbard Hamiltonian in \cref{eq_Hubbard}  belongs to the Hamiltonian class of \cref{eq_originalhamiltonian}, with
hopping terms transparent to flavor ($h_{f, f'} = -t \,\delta_{f,f'}$) and no pair creation processes ($\Delta_{f, f'} = 0$), so it can be readily defermionized.

The form of \cref{eq_Hubbard} includes the $- \frac{1}{2}$ terms in the interaction component, which are equivalent to setting a specific uniform chemical potential, to explicitly reveal the rich symmetry content of the Hubbard model. In fact, besides the obvious lattice symmetries (translation by 1 site in $x$ and $y$, $ \frac{\pi}{2}$ rotations, vertical and horizontal reflection, and compositions thereof) Hubbard dynamics exhibits two useful \emph{glocal} symmetries, where the transformation is global but comprised of separate single-site operations, both of them non-Abelian.

The first symmetry, an $SU(2)$ group, represents rotational invariance in the flavors. Since the hopping is flavor-transparent, and the interaction is based on double occupancy, the model has to be flavor-invariant. This symmetry is a Lie group, generated by the operator algebra
\begin{equation}
    \vec{S}_{\text{tot}} = \sum_{\vb{j}\in \Lambda} \vec{S}_{\vb{j}}, \quad \mbox{where} \quad
    \vec{S}_{\vb{j}} = \frac{1}{2} \sum_{f,f'} \vec{\sigma}_{f,f'} \psi^{\dagger}_{\vb{j},f} \psi_{\vb{j},f'}
\end{equation}
are the single-site flavor (spin) operators, acting non-trivially only on the singly-occupied sites.

The second symmetry, assembling total fermion number conservation and particle-hole inversion, is an $O(2)$ group, the symmetry group of the circle (rotations of a scalar angle plus one reflection). 
Despite being a continuous group, it is not Lie, but any element can be written as a rotation ${R}_{\text{tot}}(\theta) = \prod_{\vb{j}} {R}_{\vb{j}}(\theta)$, with ${R}_{\vb{j}}(\theta) = e^{i \theta (n_{\vb{j},\uparrow} + n_{\vb{j},\downarrow})}$,
eventually followed by a reflection ${F}_{\text{tot}} = \prod_{\vb{j}} {F}_{\vb{j}}$. 
It is indeed possible to write the particle-hole transformation $F$ in a way that commutes with the flavor-rotations $\vec{S}$, thus the two symmetries are indeed independent. 
Doing so yields precisely
\begin{equation} 
\label{eq_trigaugesagain}
\Bigg\{
\begin{aligned}
  R_{\vb{j}}(\theta) \, \psi_{\vb{j},f} \, R_{\vb{j}}^{\dagger}(\theta) &= e^{-i \theta} \psi_{\vb{j},f}
  \\
  F_{\vb{j}} \, \psi_{\vb{j},f} \, F_{\vb{j}}^{\dagger} &=  (-1)^{j} {\textstyle{\sum_{f'}}} \sigma^{y}_{f,f'} \psi_{\vb{j},f'}^{\dagger}
  \end{aligned}.
\end{equation}
Other symmetries, such as the pseudospin conservation \cite{LiebHubbard, HubbardReview2021},  are indeed present, but not practically useful for our numerical simulation purposes.

When defermionizing the Hubbard model for tensor network simulation, we find it convenient to enforce the \emph{vertex gauge constraint} as an exact, actual symmetry. By the prescription discussed in Ref.~\cite{LGTN}, this constraint will allow us to select a reduced canonical basis for the dressed site, made by only and all the vertex-gauge invariant states.
Conversely, the \emph{link constraint} and especially the \emph{plaquette constraint} are cumbersome to treat as exact symmetries (their local algebras do not commute) and instead are included in the Hamiltonian as penalty terms, that increase the energy of \emph{gauge} symmetry-violating sectors. 
In conclusion, we have
\begin{equation} 
\label{eq_HubbarDef_pt1}
  \begin{split}
   H'_{\text{Hub}} &= 
   -t  \sum_{f}\sum_{\vb{j},\vb*{\mu}}\qty[i \psi^{\dagger}_{\vb{j}, f} \gamma_{\vb{j},\vb*{\mu}} \gamma_{\vb{j}+\vb*{\mu},-\vb*{\mu}} \psi_{\vb{j}+\vb*{\mu}, f}]\\
    &+ U \sum_{\vb{j}} \left( n_{\vb{j}, \uparrow} -\frac{1}{2} \right)
    \left( n_{\vb{j}, \downarrow} -\frac{1}{2} \right),
  \end{split}
\end{equation}
plus a penalty
\begin{equation}
\label{eq_HubbarDef_pt2}
    \begin{split}
    H'_{\text{pen}}&= -\alpha_{b}\sum_{\vb{j},\vb*{\mu}} 
    \qty((-1)^{c^{\dagger}_{\vb{j},\vb*{\mu}}c_{\vb{j},\vb*{\mu}} + c^{\dagger}_{\vb{j}+\vb*{\mu},-\vb*{\mu}} c_{\vb{j}+\vb*{\mu},-\vb*{\mu}}} -1)\\
   &-\alpha_p
    \sum_{\square\in \Lambda} 
    \qty(\begin{array}{ccc}
    \ulcorner& \gamma_{+\vb*{\mu}_{x}} \gamma_{-\vb*{\mu}_{x}} & \urcorner\\
    \gamma_{-\vb*{\mu}_{y}} & & \gamma_{-\vb*{\mu}_{y}}  \\
    \gamma_{+\vb*{\mu}_{y}} & & \gamma_{+\vb*{\mu}_{y}} \\
    \llcorner&  \gamma_{+\vb*{\mu}_{x}} \gamma_{-\vb*{\mu}_{x}}  & \lrcorner\\
    \end{array}-1),
    \end{split}
\end{equation}
with separate penalty couplings $\alpha_p{>} 0$ and $\alpha_b {>} 0$ for the plaquette and bond violations respectively. The added $-1$ constants ensure that the correct link and plaquette symmetry sectors provide no energy contribution.
Notice that, for the mapping to be perfectly equivalent to the original model, the penalty terms of \cref{eq_HubbarDef_pt2} must represent the largest energy-scale contribution of the Hamiltonian. 
Therefore, the penalties $\alpha_p$ and $\alpha_b$ must be sufficiently larger than $t$ and $U$.
\subsection{Dressed-site Hamiltonian}
Once the vertex gauge symmetry is successfully installed, we are left with a \emph{dressed-site} vertex gauge-invariant canonical basis of dimension 32 (4 matter states and four 2-dimensional rishon modes, divided in half by the vertex constraint). Within this canonical basis, the algebra of 1-site operators is genuinely local and can be expressed in terms of the following quadratic operators
\begin{equation}
    \label{eq_defoperators}
    \begin{aligned}
        Q_{\vb{j},\vb*{\mu},f} &= \gamma_{\vb{j},\vb*{\mu}} \psi_{\vb{j},f} &
        C_{\vb{j},\vb*{\mu}_a,\vb*{\mu}_b} = \gamma_{\vb{j},\vb*{\mu}_a} \gamma_{\vb{j},\vb*{\mu}_b}\\
        W_{\vb{j},\vb*{\mu}} &= 1 - 2 c^{\dagger}_{\vb{j},\vb*{\mu}}c_{\vb{j}, \vb*{\mu}},
    \end{aligned}
\end{equation}
each one preserving local fermion parity by design.
After this re-formatting, we have a final expression for the defermionized Hubbard model, and it reads
\begin{equation} 
\label{eq_HubbardFinal}
\begin{split}
   H''_{\text{Hub}} =& 
   -t\sum_{f}\sum_{\vb{j}\in\Lambda}\sum_{\vb*{\mu}}
    \qty[i Q^{\dagger}_{\vb{j},\vb*{\mu}, f} Q_{\vb{j}+\vb*{\mu},-\vb*{\mu},f} +\text{H.c.}]\\
    &+ U \sum_{\vb{j}\in\Lambda} \qty( n_{\vb{j},\uparrow} -\frac{1}{2})
    \qty( n_{\vb{j},\downarrow} -\frac{1}{2}),
  \end{split}
\end{equation}
plus a penalty
\begin{equation} 
\label{eq_HubbarDef}
  \begin{split}
   H''_{\text{pen}} =
     &-\alpha_{b}\sum_{\vb{j}\in\Lambda}\sum_{\vb*{\mu}} 
     \qty(W_{\vb{j},\vb*{\mu}}W_{\vb{j}+\vb*{\mu},-\vb*{\mu}} -1)\\
   &-\alpha_p
    \sum_{\square \in \Lambda}\qty(\begin{matrix}
      C_{\ulcorner} & C_{\urcorner}\\
      C_{\llcorner} & C_{\lrcorner}
    \end{matrix}-1),
    \end{split}
\end{equation}
where $C$ are the \emph{corner} operators defined in \cref{eq_defoperators} and form the plaquette interaction previously defined in \cref{eq_plaquette_term} in terms of links.
Using tensor networks, we simulate the model as it exactly appears in these expressions.

\subsection{Numerical Results} \label{sec_ttnresults}
\begin{figure*}
    \centering
    \includegraphics[width=\textwidth]{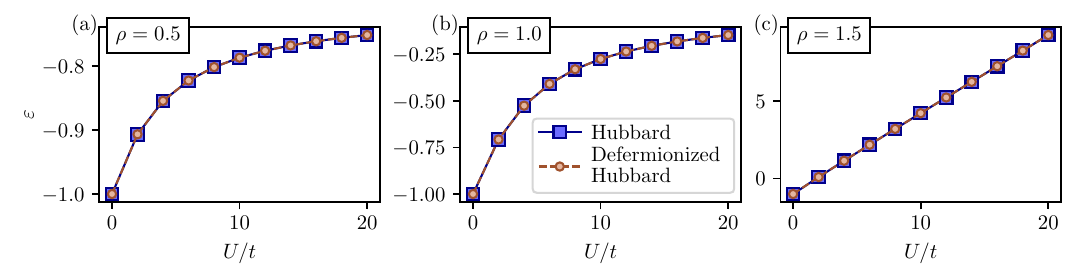}
    \caption{Exact Diagonalization comparison on a $2\times 2$ lattice between the ground state energy density $\varepsilon$ of the original 2D Hubbard model and its defermionized version as a function of $U/t$, and for three values of the particle density $\rho$: \textbf{(a)} below half-filling with $\rho=0.5$, \textbf{(b)} at half-filling with $\rho=1.0$, \textbf{(c)} above half-filling with $\rho=1.5$.}
    \label{fig_ed_comparison_2x2}
\end{figure*}
\begin{figure*}
     \centering
    \includegraphics[width=\textwidth]{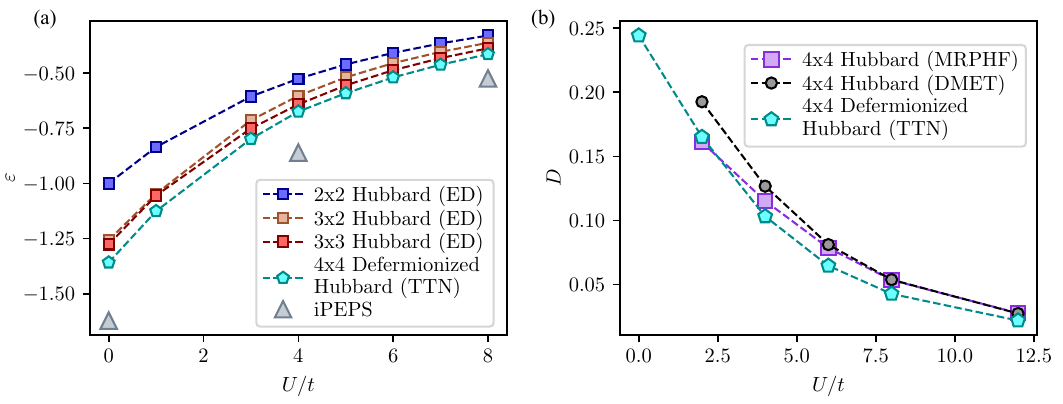}
    \caption{\textbf{(a)} Ground state energy density of the 2D Hubbard Hamiltonian at $\rho=1$ for different lattice sizes. 
    Results concerning $2\times 2$, $3\times 2$, and $3\times 3$ lattices are obtained via exact diagonalization (ED) of the original Hubbard Hamiltonian in \cref{eq_Hubbard}. 
    Energies of the $4\times 4$ lattice result from TTN simulations of the defermionized Hamiltonian in Eqs.~\eqref{eq_HubbardFinal}-\eqref{eq_HubbarDef} with bond dimension $\chi=350$ and energy penalties $\alpha_p = \alpha_b = 20$.
    \textbf{(b)} Double occupancy $D$ as a function of $U/t$ is computed with different numerical methods. 
    Data for MRPHF and DMET are taken from \cite{LeBlanc_PRX2015}.}   
     \label{fig_comparison_4x4}
\end{figure*}
In this section, we present the numerical results obtained using exact diagonalization (ED) and tree tensor networks (TTN) algorithms \cite{Cataldi2021a, PhysRevB.105.214201}. Our numerical computations concern only open boundary conditions (OBC).

To numerically check the validity of our mapping, we compare the original 2D Hubbard model of \cref{eq_Hubbard} and its defermionized version described in Eqs. \eqref{eq_HubbardFinal}-\eqref{eq_HubbarDef}. For both Hamiltonians, we perform ED on a $2 \times 2$ lattice. 

Since the two Hamiltonians do not fix a specific number of particles on the lattice, we add to them an extra term of the form $H_N = \Tilde{\mu} \qty(\sum_{\vb{j}\in \Lambda}\qty( n_{\vb{j}, \uparrow} + n_{\vb{j}, \downarrow}) - N )^2$, where $\Tilde{\mu}$ plays the role of a large penalty coefficient that increases the energy of all the states with a number of particles differing from $N$. In this way, by tuning $N$ and setting $\alpha_p =\alpha_b = 20$ and $\tilde{\mu} = 20$, we perform ED at fixed values of the particle density $\rho=N/|\Lambda|$, comparing the ground-state energy densities of the two models as a function of the ratio $U/t$. 
Notice that the chosen values of the penalty coefficients are enough to satisfy both link and plaquette constraints with a precision larger than $10^{-7}$.

As displayed in \cref{fig_ed_comparison_2x2} for three values of the particle density, i.e. $\rho\in\qty{0.5, 1.0, 1.5}$, the relative distance between the energies of the two models ($E_{H}$ for the original Hubbard model, $E_{def. H}$ for its defermionized version) $\Delta E=\abs{E_{H}-E_{def. H}}/\abs{E_{H}}<10^{-12}$ confirms the exactness of our mapping. As expected, the above half-filling particles are forced to form at least a single-site pair which determines a linear dependence of the energy with $U/t$. 

Aware of the equivalence between the two Hamiltonians, we focus on the defermionized model for larger lattice sizes by using TTN simulations. Our TTN algorithm for variational ground state search exploits the global $U(1)$ symmetry of the Hubbard model and the Krylov subspace expansion \cite{10.21468/SciPostPhysLectNotes.8}. 
Thus, the additional term $H_{N}$ is no longer needed, as the algorithm directly conserves the total number of particles by encoding the $U(1)$ symmetry sectors in the tensor networks. 

\begin{figure*}[t]
    \centering
    \includegraphics[width=\textwidth]{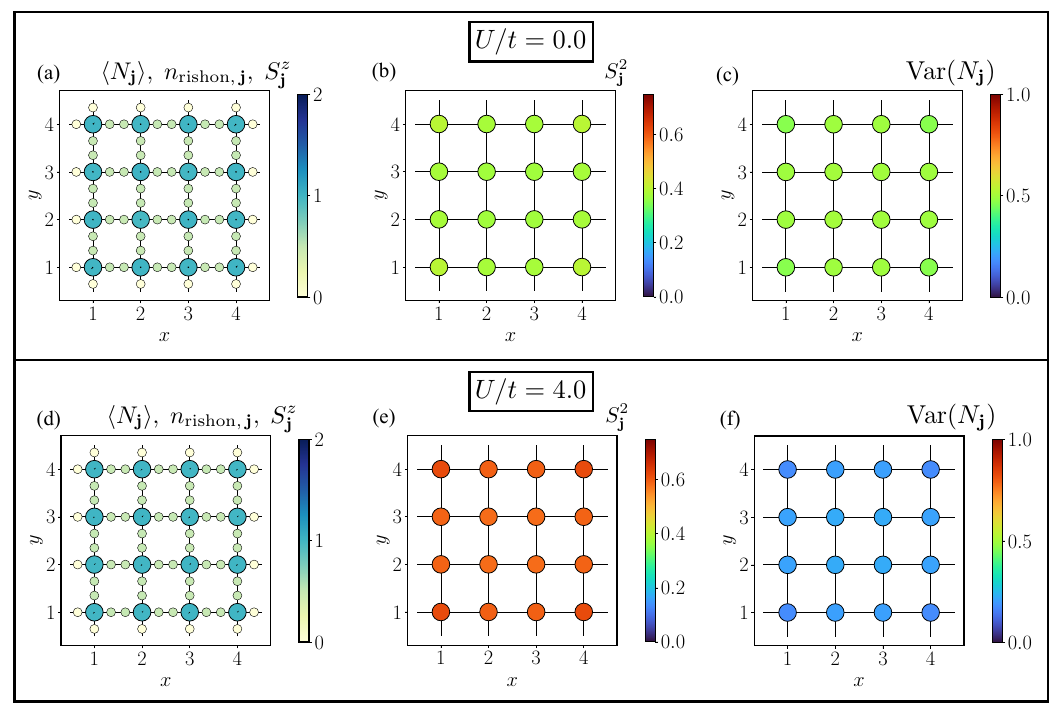}
    \caption{Ground state characterization for the defermionized Hamiltonian with $U/t=0.0, 4.0$ numerically simulated via TTNs. Computed observables: \textbf{(a)},\textbf{(d)} local configurations of fermionic matter for each lattice site $\avg{N_{\vb{j}}}$, local spin along the $z$-axis $S^z_{\vb{j}}$ represented by arrows in the center of the lattice sites (values are close to zero), rishon modes occupation on the links $ n_{\text{rishon}, \vb{j}}$; \textbf{(b)},\textbf{(e)} spin modulus squared $S^2_{\vb{j}}$; \textbf{(c)},\textbf{(f)} variance of the matter occupation number $\mathrm{Var}(N_{\vb{j}})$.}
    \label{fig_local_configurations}
\end{figure*}
Given the defermionized Hubbard Hamiltonian of Eqs. \eqref{eq_HubbardFinal}-\eqref{eq_HubbarDef}, the ground state at a specified bond dimension $\chi$ is determined by iteratively optimizing each of the tensors in TTN, gradually reducing the energy expectation value. This procedure is iterated several times to reach the desired convergence. For a detailed and exhaustive description of the algorithms, please see the technical reviews and textbooks \cite{10.21468/SciPostPhysLectNotes.8, Montangero_book}. In our TTN simulations, we use a maximum bond dimension $\chi=350$, which is sufficient to reach a relative error on the energy density of the order $10^{-6}$, ensuring the stability of our findings.

In \cref{fig_comparison_4x4}(a), we show the TTN results concerning the defermionized Hamiltonian on a $4 \times 4$ lattice at half-filling, i.e. $\rho=1$. 
We also report the corresponding energy densities obtained via ED for the original Hubbard Hamiltonian at smaller lattice sizes, and the results of the extrapolation of the infinite-size limit obtained by using iPEPS methods \cite{PhysRevB.93.045116}. 
The energies obtained with the TTN simulations for the defermionized Hamiltonian are in agreement with the overall scaling shown by the exact energies as a function of $U/t$ and the lattice sizes.

To further test the equivalence between the original Hubbard model and its defermionized version, we compute the ground-state $D\equiv 1/|\Lambda| \sum_{\vb{j}} \left < n_{\vb{j},\uparrow}n_{\vb{j},\downarrow} \right >$. In \cref{fig_comparison_4x4}(b), we show the results obtained from the TTN simulations of the defermionized model on the $4\times 4$ lattice. For the sake of comparison, we report the data of the original Hubbard Hamiltonian, obtained in Ref. \cite{LeBlanc_PRX2015} by using two independent numerical methods, i.e. the multireference projected Hartree-Fock method (MRPHF) and the Density matrix embedding theory (DMET). A good agreement and consistency of our results with the reference data is visible for all the simulated values of $U/t$. 

TTN simulations allow for an extra characterization of the ground state in terms of local observables, such as the occupation of fermionic matter and its variance 
\begin{equation}
    \begin{split}
    \avg{N_{\vb{j}}} &= \avg{n_{\vb{j},\uparrow} + n_{\vb{j}, \downarrow}}\\
    \mathrm{Var}(N_{\vb{j}}) &= \langle\qty(N_{\vb{j}})^{2}\rangle - \avg{N_{\vb{j}}}^2,
    \end{split}
\end{equation}
the local spin along the $z$-axis and its square modulus,
\begin{equation}
    \begin{split}
    S^{z}_{\vb{j}} &= \frac{1}{2}\avg{n_{\vb{j},\uparrow} - n_{\vb{j}, \downarrow}}\\
    S^2_{\vb{j}} &= \frac{3}{4}\avg{(n_{\vb{j},\uparrow} - n_{\vb{j}, \downarrow})^2},
    \end{split}
\end{equation}
but also the rishon mode occupation on lattice links $ n_{\text{rishon}, \vb{j}} = \langle c^{\dagger}_{\vb{j}, \vb*{\mu}}c_{\vb{j}, \vb*{\mu}}\rangle$.

In \cref{fig_local_configurations}, we show the results for $U/t=0$ and $U/t=4$. 
In both cases, as shown in \cref{fig_local_configurations}(a) and \cref{fig_local_configurations}(d), the fermionic occupation number $N_{\vb{j}} \sim 1$, while the local spin along $z$, represented by arrows in the center of the lattice sites, is close to zero. 
Correspondingly, all the rishon modes occupations on the links are $ n_{\text{rishon}, \vb{j}} \sim 0.5 $, as a consequence of the satisfied link and plaquette penalty constraints that encode the fermion parity. 
By looking at Figs. \ref{fig_local_configurations}(b) and \ref{fig_local_configurations}(e), we observe that the squared spin modulus $S^2_{\vb{j}}$ increases for all lattice sites by varying $U/t$ from $0.0$ to $4.0$, whereas the variance of the fermionic occupation number $\mathrm{Var}(N_{\vb{j}})$ decreases, as shown in Figs. \ref{fig_local_configurations}(c) and \ref{fig_local_configurations}(f). 

These configurations agree with the expected ground state of the original Hubbard model. Indeed, in the case of half-filling, the Hubbard model maps to the Heisenberg model \cite{Auerbach}, and the ground state mimics an antiferromagnetically long-range ordered state.

\begin{table*}[ht]
    \centering
    \begin{tabular}{l|c|c|c|c}
                    & \makecell{Qubit-fermion\\ ratio }& \makecell{Local fermion\\ parity weight} & \makecell{Hopping\\ weight} & \makecell{Stabilizer\\ weight}\\\hline\hline
       This paper                                                 & 3 & 1 & 6 & 5-6 \\
       Jordan-Wigner~\cite{nielsen2005}                           & 1 & 1 & $O(L)$ & -\\
       Bravyi-Kitaev~\cite{BRAVYI2002}                            & 1 & $O(\log_2 L^2)$ & $O(\log_2 L^2)$ & - \\
       \makecell[l]{Optimal fermion-to-qubit \\mapping~\cite{Jiang2020}}          & 1 & $O\left(\log_3{(2L^2 +1)} \right)$ & $O\left(\log_3{(2L^2 +1)} \right)$ & - \\
       Zohar-Cirac~\cite{ErezA}$^*$                                   & 3 & 1 & 5-7 & 5-6 \\
       Exact bosonization~\cite{chen2018}                         & 2 & 4 & 2-6 & 6\\
       \makecell[l]{Supercompact fermion-to-qubit\\ mapping~\cite{chen_PRXQ2023}} & 1.25 & 1-2 & 2-6 & 12\\
       
       \end{tabular}
    \caption{Comparison of different encoding mapping a single species of fermions to qubits for a $2$d square lattice of size $L\times L$. \\ * This mapping is specifically developed for lattice gauge theories, similarly to \cite{PardoPRR2023}. }
    \label{tab:encodings}
\end{table*}

\section{Digital quantum simulation}\label{sec_fqm}

The gauge `defermionization' offers a natural encoding of fermionic degrees of freedom to qubits, thus allowing for the digital quantum simulation of fermionic models on quantum computers.

Here, we break down the procedure to express the spin-$\frac{1}{2}$ Hubbard defermionized model on a square lattice in terms of qubits and Pauli operators on a generic platform, see \cref{sec_digital_fqm}, \cref{sec_digital_lc}, and \cref{sec_digital_spc}.  
This gauge-field-based encoding is genuinely local, i.e.~the lattice support of each Hamiltonian term is not increased in the encoding, and thus each Pauli weight $W$ is conserved and does not depend on the system size $N$.
However, locality comes at the cost of including auxiliary qubits representing the defermionized modes, i.e. fermionic modes and  $\mathbb{Z_2}$ gauge fields. 
There currently exists a few local mappings \cite{bravyi2022,FermitospinFrank, Whitfield_PRA2016,chen2018, Steudtner_PRA2019, Setia_PRR2019, Jiang_PRApp2019, Derby_PRB2021} and the relation between them is analyzed in \cite{chen_PRXQ2023}. 
Fermion-to-qubit mappings need resources that are usually quantified in terms of the number of qubits to simulate one fermion on average (qubit-fermion ratio), the number of qubits to express the parity of a fermionic state (parity weight), the length of the hopping operators expressed as Pauli strings and the maximum length of Pauli strings for the stabilizers.
The resources required by the gauge-field-based encoding are comparable with state-of-the-art methods: values for qubit-fermion ratio, fermion parity weight, hopping weight, and stabilizer weight are $(3, 1, 6, 6)$ respectively. Building on Table I of~\cite{chen_PRXQ2023}, we present 
a comparison of different fermionic encodings in \cref{tab:encodings}.
Within these prescriptions, we test our construction in out-of-equilibrium scenarios that will be relevant to investigate nontrivial dynamical effects.
In particular, the Hubbard model offers the opportunity to explore the dynamics of spin-like and charge-like excitations, which in one dimension manifest as distinct degrees of freedom with independent propagation velocities \cite{giamarchi2003, Kollath_PRL2005}, and have been recently observed in cold gases experiments \cite{Vijayan_Science2020, Borhdt_Rev2021} and digital quantum simulations \cite{arute2020observation}. 
In two dimensions, the dynamics of spin and charge degrees of freedom is highly nontrivial \cite{Grusdt2018} as strongly-correlated effects become relevant, still a subject of ongoing theoretical and experimental analysis with quantum simulation platforms \cite{Greiner2021, Bloch2021}.

For these reasons, we simulate the digital dynamics of spin and charge excitations over a half-filled ($\rho=1$) 2 $\times$ 4 system in the antiferromagnetic phase with $U/t=10$.
In \cref{sec_digital_methods}, we lay out the protocols to adiabatically prepare the antiferromagnetic ground state, and then inject the excitations. 
Finally, in \cref{sec_digital_scs}, we observe the corresponding out-of-equilibrium dynamics for spin and charge excitations. 
While protocols are designed for a general digital quantum computer, the presented results are obtained from tensor networks \cite{qmatchatea_0_5_2} to emulate the evolution of the quantum circuit.

\subsection{Fermion to qubits mapping of the defermionized Hubbard Hamiltonian} \label{sec_digital_fqm}
We introduce a qubit for each flavor (up $u$ and down $d$) and one for each rishon (north $n$, west $w$, east $e$, south $s$). To minimize the Pauli weight of the Hamiltonian terms, a specific ordering for the qubits composing each dressed site is defined:
\begin{align} \label{eq_even_odd_order}
    \begin{cases}
    \{u, d, w, s, e, n\} & \text{if $(-1)^{j_x+j_y}=1$ (even site)} \\
    \{d, u, s, w, n, e\} & \text{if $(-1)^{j_x+j_y}=-1$ (odd site)}. 
    \end{cases}
\end{align}
All the operators in \cref{eq_HubbardFinal} and penalties in \cref{eq_HubbarDef} are first mapped to Majorana operators and then to spin-$\frac{1}{2}$ algebra, while preserving all the commutation relations.
We recall that a hopping operator from an odd ($0$) and an even ($E$) site for the up species can be written as:
\begin{align}
     \psi_{\uparrow E}^\dagger&\gamma_{eE}\gamma_{wO}\psi_{\uparrow O} - \psi_{\uparrow E}\gamma_{eE}\gamma_{wO}\psi_{\uparrow O}^\dagger= \\
     & \gamma_{eE}\gamma_{wO}\left( \psi_{\uparrow E}^\dagger\psi_{\uparrow O} - \psi_{\uparrow E}\psi_{\uparrow O}^\dagger \right).
     \nonumber
\end{align}
Then, by defining the Majorana operators $d_{x_{\uparrow E, \uparrow O}}, d_{y_{\uparrow E, \uparrow O}}$ as:
\begin{align}
    d_{x_{\uparrow E, \uparrow O}} &= \psi_{\uparrow E, \uparrow O} + \psi_{\uparrow E, \uparrow O}^\dagger \\
    d_{y_{\uparrow E, \uparrow O}} &= i\left( \psi_{\uparrow E, \uparrow O}-\psi^\dagger_{\uparrow E, \uparrow O}\right),
\end{align}
we can trivially prove that
\begin{align}
    d_{x_{\uparrow E}}d_{y_{\uparrow O}} - d_{y_{\uparrow E}}d_{x_{ \uparrow O}} = 2i\left( \psi_{\uparrow E}^\dagger\psi_{\uparrow O} - \psi_{\uparrow E}\psi_{\uparrow O}^\dagger \right).
\end{align}
We can thus express the hopping in terms of the Majorana operators.
Following the order defined in \cref{eq_even_odd_order}, the operators acting on an even site $j$ can be written as:
\begin{align}\label{eq_qub_map}
    d_{x_{\vb{j},\uparrow }} = XZZZZZ, \quad&  d_{x_{\vb{j},\downarrow }} = IXZZZZ,\\ \label{eq_qub_map_1}
    d_{y_{\vb{j},\uparrow }} = YZZZZZ, \quad&  d_{y_{\vb{j},\downarrow }} = IYZZZZ,\\ \label{eq_qub_map_2}
    \gamma_{\vb{j},r} = I_uI_d\bigotimes_{k=w}^{r-1}I_k &\otimes X_r \bigotimes_{k=r+1}^{n}Z_k,
\end{align}
where $X,Y,Z$ are Pauli matrices and $I$ is the identity. In the two equations above we dropped the index of species and rishons for clarity.
The string of $Z$ operators applies to all the qubits forming a site. 
The index $r$ runs over the different rishon species of a site, as defined in \cref{eq_even_odd_order}. 
The mapping for odd sites can be obtained analogously using Eq.~\eqref{eq_even_odd_order}, and is equivalent to swapping the first two operators and using the odd-ordered sites.

Now, we compute the explicit form of the Hamiltonian terms, hopping and on-site interaction, in terms of Pauli operators. 
Following the notation defined in \cref{eq_even_odd_order}, the horizontal hopping operator from an odd to an even site for flavor $up$ reads:
\begin{equation}
\resizebox{\columnwidth}{!}{
   \begin{tabular}{c|cccccc|cccccc}
   & u & d & w & s & e & n & d & u & s & w & n & e \\ \hline
    $d_{x_{\uparrow E}}$&X& Z & Z & Z & Z & Z & I & I & I & I & I & I \\
    $\gamma_{eE}$& I & I & I & I & X & Z & I & I & I & I & I & I \\
    $\gamma_{wO}$& I & I & I & I & I & I & I & I & I & X & Z & Z \\
    $d_{y_{\uparrow O}}$& I & I & I & I & I & I & I &Y& Z & Z & Z & Z \\ \hline
    hopping    &X& Z & Z & Z & iY &I & I &Y& Z  & -iY & I & I 
\end{tabular}
}
\end{equation}
Matrix multiplication goes from bottom to top. By computing also the hermitian conjugate, the hopping term results:
\begin{align}\label{eq_hopping_qubs}
    d_{x_{\uparrow E}}&\gamma_{eE}\gamma_{wO}d_{y_{\uparrow O}} - d_{y_{\uparrow E}}\gamma_{eE}\gamma_{wO}d_{x_{\uparrow O}}= \\ 
   &\left(XZZZYIIYZYII-YZZZYIIXZYII \right). \nonumber
\end{align}
Terms for vertical hopping or hoppings involving different species can be derived analogously.

Similarly, we derive the on-site interaction term. 
The number operator acting on site $\vb{j}$ can be written as $n_{\vb{j},\sigma}=\frac{1}{2}\left(1-Z_{\vb{j},\sigma}\right)$, then the on-site interaction reads:
\begin{align}\label{eq_onsite_qubs}
    \left(n_{\vb{j},\uparrow}-\frac{1}{2}\right)\left(n_{\vb{j},\downarrow}-\frac{1}{2}\right)=\frac{1}{4}Z_{\vb{j},\uparrow}Z_{\vb{j}, \downarrow}.
\end{align}

Once we obtain the Hamiltonian terms described as Pauli strings, we decompose the relative time propagator in single and two-qubit gates following Ref.~\cite{Whitfield_PRA2016}. 
A detailed description of the mapping from the time propagator to the quantum circuit can be found in Appendix~\ref{app:two_qubits}.

\subsection{Vertex and plaquette constraints}\label{sec_digital_spc}
Following the prescription of the gauge defermionization, we have to constrain the state to a specific subspace of the Hilbert space, a condition set by the first two lines of \cref{eq_trigauges}.
Each symmetry can be cast into a language that directly translates into a quantum circuit scenario and is equivalent to a stabilizer of a quantum error correcting code; to respect the gauge symmetry, the state is restrained to the $+1$ eigenstate of the stabilizers~\cite{gottesman1997}.
The Pauli representation of the stabilizers can be obtained following the mapping presented in \cref{eq_qub_map}. As expected, all the stabilizers commute with the Hamiltonian. 

The vertex stabilizer term reads from the first line of \cref{eq_trigauges}. As the exponentials are counting the parity of a given species/rishon, these operators correspond to $Z$ operators in the qubit language. Thus, the stabilizer acting on the lattice vertex $\vb{v}$ results in
\begin{align}
    S_{\vb{v}} = \bigotimes_{f=u, d} Z_f \bigotimes_{\mu\in \vb{v}}Z_{\vb{v}+\vb*{\mu}},
\end{align}
where $\mu$ runs on all rishons connected to the vertex $\vb{v}$. This stabilizer must be applied to each vertex $\vb{v}$ in the system.

The plaquette stabilizer, similar to the one in \cite{PardoPRR2023}, reads from the second row in \cref{eq_trigauges}. 
As the product always runs on four terms, the imaginary unit $i$ drops. 
Following the representation of the $\gamma$ operators reported in \cref{eq_qub_map_2}, the final form of each plaquette term respects the qubits ordering of the dressed site. 
Thus, the form of an even plaquette stabilizer is different from an odd one, where a plaquette is said even (odd) if the lower-left corner of the plaquette lies on an even (odd) site. We report the explicit computation for an even plaquette in \cref{tab:plaquette_stabilizer}.

Finally, we report the size of the Hilbert space to perform a digital quantum simulation of the defermionized Hubbard model over a $x\times y$ grid, and compare it to the size of the accessible subspace:
\begin{align}
    \text{dim}(\mathcal{H}) &= 2^{ 2xy+x(y-1)+y(x-1) }=2^{4xy-x-y} \\
    \text{dim}(\mathcal{H}_{eff}) &= 2^{ 2xy+x(y-1)+y(x-1)-(x-1)(y-1)-xy }\nonumber \\ &=2^{2xy-1}.
\end{align}
We notice that constraining the system to the physical subspace quadratically decreases the dimension of the Hilbert space, showing that we are working with a highly redundant space.

\begin{table*}[t]
    \centering
    \resizebox{\textwidth}{!}{
    \begin{tabular}{c|cccccc|cccccc|cccccc|cccccc}
                 & u & d & w & s & e & n & d & u & s & w & n & e & d & u & s & w & n & e & u & d & w & s & e & n \\ \hline
  $\gamma_{00e}$ & I & I & I & I & X & Z & I & I & I & I & I & I & I & I & I & I & I & I & I & I & I & I & I & I \\
  $\gamma_{00n}$ & I & I & I & I & I & X & I & I & I & I & I & I & I & I & I & I & I & I & I & I & I & I & I & I \\
  $\gamma_{01s}$ & I & I & I & I & I & I & I & I & X & Z & Z & Z & I & I & I & I & I & I & I & I & I & I & I & I \\
  $\gamma_{01e}$ & I & I & I & I & I & I & I & I & I & I & I & X & I & I & I & I & I & I & I & I & I & I & I & I\\
  $\gamma_{10w}$ & I & I & I & I & I & I & I & I & I & I & I & I & I & I & I & X & Z & Z  & I & I & I & I & I & I\\
  $\gamma_{10n}$ & I & I & I & I & I & I & I & I & I & I & I & I & I & I & I & I & X & Z & I & I & I & I & I & I\\
  $\gamma_{11w}$ & I & I & I & I & I & I & I & I & I & I & I & I & I & I & I & I & I & I & I & I & X & Z & Z & Z\\
  $\gamma_{11s}$ & I & I & I & I & I & I & I & I & I & I & I & I & I & I & I & I & I & I & I & I & I & X & Z & Z\\ \hline
    stabilizer & I & I & I & I & X & iY & I & I & X & Z & Z & iY & I & I & I & X & iY & I & I & I & X & iY & I & I\\
\end{tabular}
}
    \caption{Explicit definition of the stabilizer for an even plaquette. This calculation does not take into account the link symmetry: thus it is necessary to do that further computation to obtain the final expression of the qubit mapping.}
    \label{tab:plaquette_stabilizer}
\end{table*}

\subsection{Link constraint}\label{sec_digital_lc}
\begin{figure}[ht]
    \centering
    \includegraphics[width=1\columnwidth]{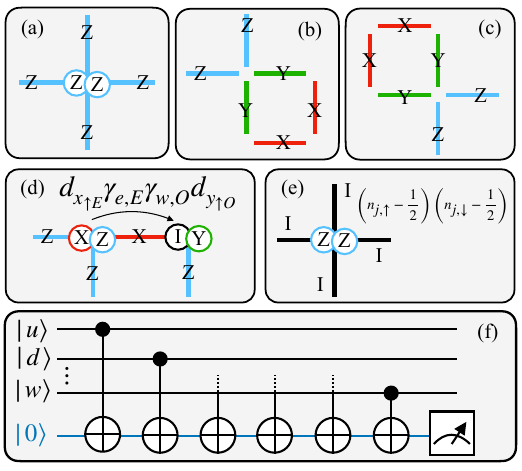}
    \caption{Graphical representation of the Hamiltonian terms in the defermionized Hubbard model and of the stabilizers for installing the  vertex, link, and plaquette symmetries. 
    These terms result from applying \cref{eq_from2_to1_rishon}, where the states of the two rishons satisfying the link constraint are mapped to a single qubit for each link. In \textbf{(a)}, we depict the vertex stabilizer, while \textbf{(b)} corresponds to the even plaquette stabilizer, and \textbf{(c)} represents the odd plaquette stabilizer. Figures \textbf{(d)} and \textbf{(e)} illustrate the hopping term and the onsite interaction respectively. As dictated by \cref{eq_even_odd_order}, in figure \textbf{(d)}, the $u$ qubit is the first on the even site and the second on the odd site. In \textbf{(f)} we report the quantum
    circuit used to measure the stabilizer in \textbf{(a)} using CNOT gates and a projective measurement on an ancilla qubit (depicted in blue). The qubits $\ket{n}, \ket{e}, \ket{s}$ are not reported in the figure, even if their CNOT gate is present with a dashed line.
    }
    \label{fig_stabilizers}
\end{figure}

The state of the two rishons sharing the same link is also subject to a constraint set by the third line of \cref{eq_trigauges}. The only qubits states fulfilling this constraint are $\ket{00}$ and $\ket{11}$. The effective Hilbert space for a pair of rishons is $2$-dimensional; for this reason, both rishons can be described with a single qubit, by mapping $\ket{00}\rightarrow \ket{0}, \ket{11}\rightarrow\ket{1}$. After this procedure, we need to project all the operators acting on the two rishons on the new subspace. Given an operator $A_{r_1r_2}$ we can write the new operator living on the two-dimensional space as:
\begin{align}\label{eq_from2_to1_rishon}
    A_{\overline{r}} = \text{Tr}_{\ket{01},\ket{10}} A_{r_1r_2},
\end{align}
where we trace out the states violating the constraint. 
We report the new form of some operators of interest as an example: $ZZ\rightarrow I$, $XX\rightarrow X$. This way, each dressed site shares a rishon with its nearest neighbors, and thus there is no longer a clear separation between the dressed sites.
In \cref{fig_stabilizers}, we report the final form of the stabilizers and the Hamiltonian terms, once the link constraint is applied.

\subsection{Preparation of spin and charge excitations}\label{sec_digital_methods}
Here, we lay out the protocols to adiabatically prepare the antiferromagnetic ground state, and to inject spin- and charge-excitations.

\paragraph{Adiabatic ground state preparation} 
The adiabatic state preparation targets the ground state of the defermionized Hubbard model over a $4\times 2$ lattice, at half-filling ($\rho=1$) with $t=0.1$ and $U=1$. 
The adiabatic evolution starts from the ground state of \cref{eq_HubbardFinal} with $t=0$, here renamed $H_0=H''_{Hub}(t=0)$; the hopping terms are slowly switched on to reach $H_1=H''_{Hub}$. Then, the time-dependent Hamiltonian reads:
\begin{align}
    H &= (1-\beta)H_0+\beta H_1,
\end{align}
where $\beta \in [0,1]$ is the adiabatic parameter that increases linearly for $100$ steps; for each $\beta$, the system evolves for ten time-steps $d\tau$ to ensure a smooth convergence, for a total of $1000$ evolution steps. 
The real-time evolution is simulated by decomposing the evolution operator via a first-order Trotterization with time step $dt=d\tau=0.01$.

Since the initial state of $H_0$ is hugely degenerate, we select as initial state for the adiabatic process is the ground state of $H_0$ at half-filling ($\rho=1$) and respects the spin-flip symmetry. 
While neglecting the rishons' state, we consider only the state of the matter qubits ($u$ and $d$) for each dressed site over the $4\times 2$ lattice, where $\ket{10}=\ket{\uparrow}$ and $\ket{01}=\ket{\downarrow}$. 
It results in
\begin{align}\label{eq_adiabatic_initial_state}
    \frac{1}{\sqrt{2}}\left(\left|
    \begin{matrix}
    \uparrow & \downarrow & \uparrow & \downarrow \\
    \downarrow & \uparrow & \downarrow & \uparrow
    \end{matrix}
    \right\rangle +
    \left|
    \begin{matrix}
    \downarrow & \uparrow & \downarrow & \uparrow \\
    \uparrow & \downarrow & \uparrow & \downarrow
    \end{matrix}
    \right\rangle\right).
\end{align}
Notice that the state in \cref{eq_adiabatic_initial_state} is a GHZ state, thus it can be easily prepared using only Hadamard, NOT, and controlled-NOT gates.
Finally, the rishons' state is uniquely determined once the stabilizers are measured, after which we apply a conditional operation to ensure the state lies in the correct symmetry sector. For a depiction of the stabilizers see \cref{fig_stabilizers}, while for an example of the measurement of the stabilizer depicted in \cref{fig_stabilizers}(a) see \cref{fig_stabilizers}(f). This measurement is performed through a projective measurement of an ancilla qubit.

At the end of the adiabatic evolution, we checked that local spin projection $ S^z_{\vb{j}}$ and local charge $ N_{\vb{j}}$ are stationary up to oscillations compatible with the cut of the singular values in the tensor network emulator ($10^{-8}$). 
For a discussion about convergence see \cref{app_convergence}.

While an adiabatic state preparation is performed in this case, we do not think this approach to be scalable on real quantum devices. However, there are multiple possibilities for preparing the initial state with shallower circuits. If a tensor network state can efficiently represent the target state, we can variationally map the state to a quantum circuit~\cite{Rudolph2024}. Another possibility is to implement a Quantum Approximate Optimization Algorithm (QAOA), which can encode the Hamiltonian interactions in the quantum circuit structure~\cite{wauters2020}. Both approaches have the advantage of being able to tune the fidelity of the algorithm with the depth of the quantum circuit.

\begin{figure*}
    \centering
    \includegraphics[width=1\textwidth]{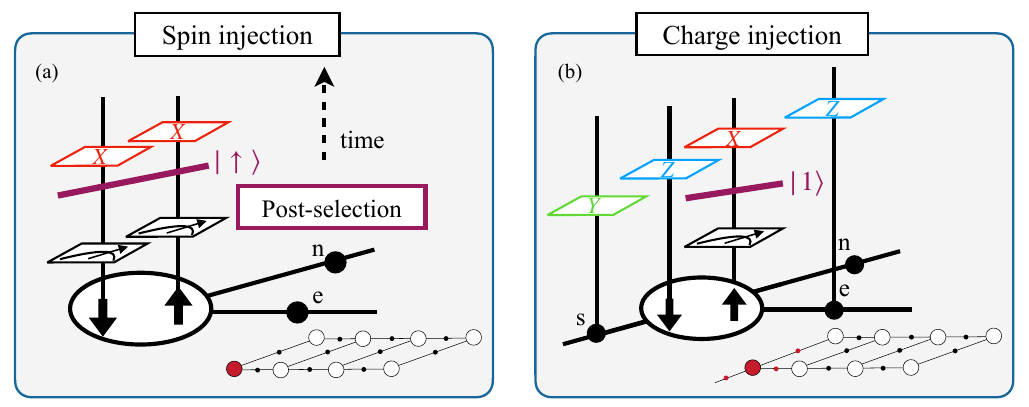}
    \caption{Schematic of injecting spin- and charge-excitations on the $(0,0)$ corner. Both figures depict the qubit lattice, highlighting up and down qubits within the matter site. Rishons are represented as black dots and labeled with respect to the corner. In the bottom right of each subfigure, we report the entire lattice, highlighting in red the site and rishons involved in the operation. All applied operations commute with the stabilizers. \textbf{(a)} \textit{Spin excitation:} To inject the spin excitation, we initially measure the qubit states for up and down, post-selecting only the $\ket{01}=\ket{\uparrow}$ state. Then, the $XX$ operator flips the spin state. \textbf{(b)} \textit{Charge excitation:} Injecting the charge excitation involves measuring the up qubit and post-selecting the state $\ket{1}$. We then apply the Pauli operators as depicted in the figure, equivalent to causing the matter qubit $u$ to jump outside the lattice. To achieve this, an additional rishon (the $s$ rishon in this case) must be added.}
    \label{fig_excitation_sketch}
\end{figure*}

\paragraph{Excitations} 
Spin- and charge-excitations are injected in the ground state of $H_1$ using local operations and classical communication (LOCC)s: this procedure is, in principle, completely reproducible on state-of-the-art quantum computers. 
Both protocols preserve the symmetries of the system, namely every constraint previously defined is satisfied. In \cref{fig_excitation_sketch}, we provide a graphical representation of the two methods.

\textbf{Spin excitation}. 
Without losing any generality, the spin excitation is created on the site $(0, 0)$. 
Recalling that qubits $u,d$  represent up and down flavors, we choose a general superposition of the two states; after a projective measurement on both qubits, only the flavor up is post-selected, i.e. the qubits state $\ket{10}=\ket{\uparrow}$.
Then, a spin excitation is introduced by flipping the spin state:
\begin{align}
    \ket{\uparrow} &\longrightarrow \ket{\downarrow},&
    \ket{1_u0_d} &\xrightarrow{XX} \nonumber\ket{0_u1_d}.
\end{align}
This operation corresponds to applying the operator $XX$ on the qubits $u, d$. Notice that $XX$ commutes with all the stabilizers, since it shares support only with the vertex stabilizer in $(0,0)$, and they commute.

\textbf{Charge excitation}. In principle, the charge excitation can be created anywhere on the lattice. However, for improved efficiency, we restrict the protocol to the sites along the border and in particular to the site $(0, 0)$. A projective measurement is performed on the $u$ qubit, then post-selecting the state $\ket{1}$.
Being left with a single charge on the site, the excitation is obtained by removing that very charge. However, this operation does not commute with the vertex stabilizer in $(0,0)$, thus breaking that constraint. To preserve Gauss' law we introduce an additional qubit, labeled as an extra rishon of site $(0,0)$, in this case, $s$ (or $w$). The excitation is implemented by flipping the state of the qubits $u, s$: 
\begin{align}
    \ket{1_u0_s} &\stackrel{XX}{\longrightarrow} \nonumber \ket{0_u1_s}.
\end{align}
This procedure, however, does not commute with the stabilizers, and it would bring the state outside the subspace of the physical states. To avoid this issue, we implement
this operation by applying the hopping operator $i\gamma_{\vb{j},s}\psi_{\vb{j},\uparrow}$ with $\vb{j}=(0,0)$, which makes the particle in $u$ hop through $s$ outside the lattice. The operator reads $g_{charge}=X_uZ_dZ_wY_s$, and it is equivalent to the bit-flip just discussed.
We stress that qubit $d$ is not projected during this protocol.

\subsection{Propagation of spin and charge excitations}\label{sec_digital_scs}
By employing the local encoding, we simulate the dynamics of spin and charge excitations over a half-filled ($\rho=1$) $2\times4$ system in the antiferromagnetic phase $U/t=10$. 
To this aim, we use quantum matcha TEA~\cite{qmatchatea_0_5_2}, an emulator of quantum circuits based on MPS. All the simulations converge with bond dimension $\chi=1024$ - below the maximum achievable for $27$ qubits, where the maximum bond dimension is bound by $\chi_{\mathrm{max}}=8192$ - and Trotter step $dt=0.01$. 
In this regime with $U\gg t$, the Hubbard model dynamics is effectively described by the $t-J$ model \cite{fradkin2013}, where high-energy doublon states are perturbatively removed and lead to an antiferromagnetic spin-exchange coupling $J=2t^2/U$ of the Heisenberg type. 
In this regime, we therefore expect to observe slower spin dynamics governed by the $J$ coupling and faster hole dynamics governed by $t$.

\begin{figure*}[t]
    \centering
    \includegraphics[width=\textwidth]{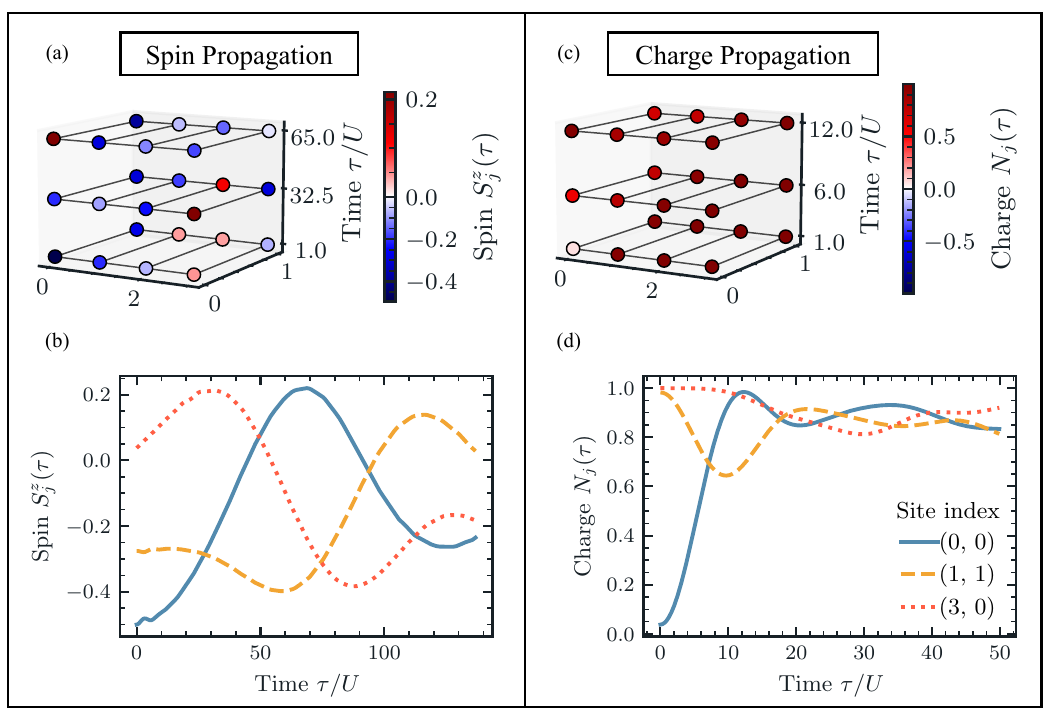}
    \caption{Digital quantum simulation of spin-charge dynamics in the $t-J$ model limit: 
    (a) Propagation of the spin $S^z_j$ profile on each site at different times in the evolution $\tau/U=1, 32.5, 65$. 
    (b) Spin profile for selected sites on the lattice.
    (c) Propagation of the charge $N_j$ profile on each site at different times in the evolution $\tau/U=1, 6, 12$. 
    (d) Charge profile for selected sites on the lattice.
    Notice that different timescales for spin and charge dynamics occur as $t\gg J$.
    }
    \label{fig_excitations}
\end{figure*}

The first column of \cref{fig_excitations} shows the evolution of the spin excitation.
By measuring the local spin along the $z$ direction $S_{\vb{j}}^{z}(\tau)$, one can monitor in time
 the deviations from the initial condition, i.e. the excitation injection:
\begin{align}
    S_{\vb{j}}^{z}(\tau)=\frac{1}{2}\left(\langle n_{\vb{j}, \uparrow}\rangle(\tau) - \langle n_{\vb{j}, \downarrow}\rangle(\tau)\right).
\end{align}
This assessment requires only the expectation values of local observables, since
\begin{align}\label{eq_n_to_qubits}
    \langle n_{\vb{j},  \gfrac{\uparrow}{\downarrow}}\rangle &= \frac{1-\langle Z_{\vb{j}, \gfrac{u}{d}} \rangle}{2}, &
    \text{where} &&\langle Z_{\vb{j}, \gfrac{u}{d}}  \rangle
\end{align}
is the expectation value of the $u(d)$ qubits over the $Z$ basis.
The second column of \cref{fig_excitations} shows the evolution of the injected charge excitation.
Similarly to the previous case, we measure the local charge $N_{\vb{j}}(\tau)$: 
\begin{align}
    N_{\vb{j}}(\tau) = \langle n_{\vb{j}, \uparrow}\rangle(\tau) + \langle n_{\vb{j}, \downarrow}\rangle(\tau),
\end{align}
where $\tau$ is the time dependence.
The numerical findings are in agreement with the $t-J$ model description, namely show much faster hole dynamics ($\tau_h \sim 1/t$) and slower spin dynamics ($\tau_s \sim 1/J$). 
This is also quantitatively confirmed for the parameters chosen in these simulations by the corresponding monitored observables. 
Considering the injection site $\mathbf j = (0,0)$, the first peak in the charge sector, $N_j(\tau)$, occurs at $\tau_h\approx 12/U = 1.2 /t$, whereas the first peak in the spin sector, $S^z_j(\tau)$, occurs at $\tau_s \approx 65/U = 6.5/t$. 
The corresponding ratio $\tau_s/\tau_h = 5.4$ is close to the ratio $t/J=U/2t = 5$, as expected.
Notice that this analysis only provides evidence of the short timescale propagation of the excitations and is additionally affected by the small system size.
While in 1D spin and charge would manifest as independent degrees of freedom also at longer times due to the well-known spin-charge separation phenomenon, this will not be the case in two-dimensional lattices.
Spin and charge would indeed display strongly-correlated dynamics at longer times, as a consequence of spinon-holon coupling (see, for example, Refs.~\cite{Grusdt2018, Borhdt_Rev2021} and references therein).
 
\section{Conclusions}\label{sec_conclusion}
We have generalized a technique for local fermion encoding to any 2D lattice configurations eliminating the fermionic degrees of freedom by absorbing them into an auxiliary $\mathbb{Z}_2$ gauge ﬁeld (gauge defermionization). 
We have successfully tested this method against the 2D spin-$\frac{1}{2}$ Hubbard model. 
The ground state properties have been computed for varying particle densities utilizing both tree-tensor network ansatz and exact diagonalization methods. 
Here, we have observed the expected transition from the liquid to the anti-ferromagnetically ordered phase, accessing lattice sizes up to $4\times4$. 
Furthermore, gauge defermionization introduces a fermion-to-qubit mapping with comparable resource requirements to state-of-the-art encodings. 
We have shown that this mapping offers a scalable pathway for the digital quantum simulation of fermionic theories when the available fault-tolerant quantum computers~\cite{bluvstein2024,dasilva2024} will be scaled. 
For the defermionized 2D Hubbard Hamiltonian, hopping terms and gauge constraints, here included as stabilizers, result in a maximum Pauli weight of 6. 
We have then demonstrated the feasibility of this approach by simulating the digital out-of-equilibrium dynamics of the defermionized 2D Hubbard Hamiltonian over a $4\times2$ lattice in the antiferromagnetic phase. 
Our protocol entails the adiabatic preparation of the ground state (half-filling, $U/t=10$), the injection of a charge (spin)-excitation in the system, and the time evolution of the perturbed state. 
Finally, we have observed a faster propagation for the charge excitation compared to the spin one as expected from the low-energy description based on the $t-J$ model.

Our collected results show numerical evidence that the local encoding is a scalable and feasible pathway toward the digital quantum simulation of fermionic lattice theories. We have highlighted that is indeed feasible to provide physically relevant results with the current technology of digital quantum processing platforms.

Tensor networks also showed positive results, although specifically the tree tensor network ansatz state manifests some limitations to accommodate the area-law of entanglement introduced by the auxiliary, resonant gauge fields. 
We expect tensor network geometries capable of capturing a wider entanglement distribution, such as ATTN \cite{ATTN} or iPEPS \cite{iPEPSalgo}, to yield even better results.

\medskip

\emph{Code and data availability $-$}
The code to map the fermionic Hamiltonian to a defermionized one and run the digital quantum simulation
is available at~\cite{code}. The engine of the simulations are distributed through the quantum tea leaves~\cite{qtealeaves_0_5_12}
and quantum matcha tea~\cite{qmatchatea_0_5_2} python packages of \emph{Quantum TEA}.

\medskip
\emph{Acknowledgments $-$}
We thank E. Ercolessi for the helpful discussions. Authors acknowledge financial support:
from the Italian Ministry of University and Research (MUR) via
PRIN2017 and PRIN2022 project TANQU, and the INFN project QUANTUM;
from the German Federal Ministry of Education and Research (BMBF) via the project QRydDemo;
from the European Union
via QuantERA projects QuantHEP and T-NiSQ,
the H2020 Quantum Flagship project PASQuanS2, Horizon Europe project EuRyQa, H2020 project TEXTAROSSA,
and via the NextGenerationEU PNRR project CN00000013 $-$ Italian Research Center on HPC, Big Data, and Quantum Computing.
We acknowledge computational resources by the Cloud Veneto, CINECA, the BwUniCluster, the University of Padova Strategic Research Infrastructure Grant 2017: “CAPRI: Calcolo ad Alprestazionioni per la Ricerca e l’Innovazione”, and the WCRI-Quantum Computing and Simulation Center of Padova University.

\appendix

\section{Mapping the time propagator to a quantum circuit}\label{app:two_qubits}
\begin{figure}
    \centering
    \includegraphics[width=\columnwidth]{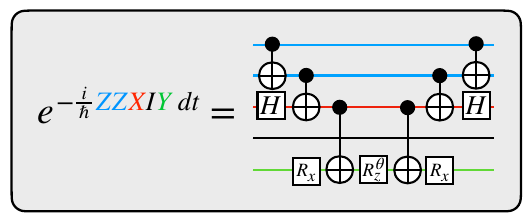}
    \caption{Mapping of the time propagator of a generic Pauli string to a quantum circuit using only single and two-qubit gates.}
    \label{fig:ham_to_gate}
\end{figure}

To simulate the dynamics of our system on a quantum computer it is important to map the evolution operator to a quantum circuit using single and two-qubit gates since those are the available operations. First, we Trotterize the evolution treating each Hamiltonian term separately. Then, it has been shown that the propagator $e^{-\frac{i}{\hbar} J H_i dt}$ of a $H_i =ZZ\dots Z$ interaction with strength $J$ can be compiled by combining a cascade of CNOTs gates and a rotation along the $z$ axis~\cite{Whitfield_PRA2016}:
\begin{align}
    R_z^\theta = \begin{pmatrix}
    1 & 0 \\
    0 & e^{i\theta}
    \end{pmatrix}, \quad \theta = Jdt.
\end{align}
In general, we can treat an arbitrary Pauli string by moving to the $Z$ basis along the CNOTs cascade. Practically, we apply the following single-qubit basis change gate before and after the CNOT, respectively for the $X$ and $Y$ Pauli matrices:
\begin{align}
     H = \frac{1}{\sqrt{2}}\begin{pmatrix}
    1 & 1 \\
    1 & -1
    \end{pmatrix},\quad
     R_x\left(-\frac{\pi}{2}\right) = \frac{1}{\sqrt{2}}\begin{pmatrix}
    1 & i \\
    i & 1
    \end{pmatrix}.
\end{align}
In Figure~\ref{fig:ham_to_gate} we report the form of the CNOTs cascade with the basis change, showing an example for a general Pauli string.

\section{Convergence checks}\label{app_convergence}
\begin{figure}
    \centering
    \includegraphics{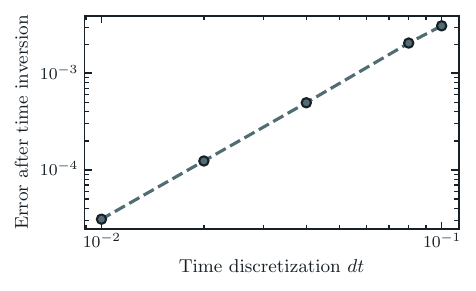}
    \caption{Trotter error as a function of the Trotter step in a log-log scale.}
    \label{fig_trotter_error}
\end{figure}

We perform different checks to ensure that the results obtained from the time evolution of the MPS are meaningful. First, one should ensure that the Trotterization step is small enough. We perform a time evolution up to time  $t/U=20$ and then invert the time, letting the system evolve back. If we are not making any Trotter errors, we should go back to the very same state. We estimate the error as the absolute value of the difference of the observable $\langle n_{\uparrow,j}n_{\downarrow,j}\rangle$ at $\tau=0$ and at the end of the evolution.
In \cref{fig_trotter_error}, we show the scaling of the error for different Trotter steps. We can safely state that the chosen Trotter step of $dt=10^{-2}$ is small enough to obtain reasonable results.

The second source of error is the state truncation due to the finite bond dimension of the system, which is set at $\chi=2^{10}=1024$. 
This bond dimension is lower than the maximum bond dimension that a system of $27$ would require, i.e. $\chi_{\mathrm{max}}=2^{13}=8192$. 
However, we can monitor the approximation we make~\cite{jaschke2022}: after each two-qubits gate, we perform at most an error $E=\sum_{i}s_i^2$, where $s_i$ are the singular values truncated by the MPS algorithm. 
A lower bound of the fidelity $\mathcal{F}$ of the final state of the simulation is:
\begin{align}
    \mathcal{F}=\prod_k (1-E_k),
\end{align}
where the index $k$ runs over all the gates of the quantum circuit implementing the time evolution. 
We obtained infidelity of $1-\mathcal{F}_s=4\cdot 10^{-7}$ for the evolution after the application of the spin excitation, and $1-\mathcal{F}_c=2\cdot 10^{-9}$ for the evolution after the insertion of the charge excitation. We can thus state that all sources of numerical errors are under control.

\section{Initial state for the adiabatic evolution}
In \cref{sec_digital_methods}, we discussed the adiabatic preparation of the ground state of the defermionized Hubbard Hamiltonian for a $4\times 2$ lattice at half filling. 
However, we can still choose the initial state for such a procedure, as long as it is a ground state of $H_0$. 
It is thus convenient to choose an easy-to-prepare state that maximizes the overlap with the expected final state. The following discussion will ignore the states of the rishons, which are always considered to be in the correct state to ensure that all the constraints are satisfied. 
Since we are interested in the strongly repulsive regime of $U/t=10$, $U>0$, we can expect to find only one particle in a given site, i.e., the up $u$ or down $d$ spin species. This is equivalent to saying that the only qubit states we allow are $\ket{1_u0_d}, \ket{0_u1_d}$. 
We also notice that any combination of states that satisfy this condition is a ground state of the defermionized Hubbard Hamiltonian in \cref{eq_HubbardFinal} with $t=0$. 
First, we consider that, once the hopping is turned on, some configurations will be favored: those that have a checkerboard configuration of sites, where we alternate the up and down species, i.e., if the site $\vb{j}$ is in the state $\ket{1_u0_d}$ all its neighboring sites will be in $\ket{0_u 1_d}$. 
We finally prepare a superposition of the checkerboard states, such that the resulting state is symmetric under a global spin flip, i.e., a flipping of the $u, d$ qubits since we expect this to be a symmetry of the Hamiltonian.
The protocol to prepare this state on a quantum computer is as follows:
\begin{enumerate}
    \item Prepare the checkerboard state;
    \item Stabilize the state under the global spin flip stabilizer, i.e. an $X$ operator acting on all the $u, d$ qubits;
    \item Stabilize the state w.r.t. all the local stabilizer defined in \cref{sec_fqm}.
\end{enumerate}
This state is an optimal choice of the initial state of the simulation.

\bibliographystyle{quantum}
\bibliography{refs}

\begin{thebibliography}{10}

\bibitem{fradkin2013}
Eduardo Fradkin.
\newblock ``Field theories of condensed matter physics''.
\newblock
  \href{https://dx.doi.org/https://doi.org/10.1017/CBO9781139015509}{Cambridge
  University Press}. ~(2013).

\bibitem{Auerbach}
Assa Auerbach.
\newblock ``Interacting electrons and quantum magnetism''.
\newblock
  \href{https://dx.doi.org/https://doi.org/10.1007/978-1-4612-0869-3}{Springer
  New York, NY}. ~(1994).

\bibitem{giamarchi2003}
Thierry Giamarchi.
\newblock ``Quantum physics in one dimension''.
\newblock
  \href{https://dx.doi.org/https://doi.org/10.1093/acprof:oso/9780198525004.001.0001}{Volume
  121}.
\newblock Clarendon press. ~(2003).

\bibitem{Lee_RevModPhys2006}
Patrick~A. Lee, Naoto Nagaosa, and Xiao-Gang Wen.
\newblock ``Doping a mott insulator: Physics of high-temperature
  superconductivity''.
\newblock \href{https://dx.doi.org/10.1103/RevModPhys.78.17}{Rev. Mod. Phys.
  {\bf 78}, 17--85}~(2006).

\bibitem{Hartke2023a}
Thomas Hartke, Botond Oreg, Carter Turnbaugh, Ningyuan Jia, and Martin
  Zwierlein.
\newblock ``Direct observation of nonlocal fermion pairing in an attractive
  {{Fermi-Hubbard}} gas''.
\newblock \href{https://dx.doi.org/10.1126/science.ade4245}{Science {\bf 381},
  82--86}~(2023).

\bibitem{FQHCollection}
Bertrand~I Halperin and Jainendra~K Jain.
\newblock ``Fractional quantum hall effects: New developments''.
\newblock \href{https://dx.doi.org/10.1142/11751}{World Scientific}. ~(2020).

\bibitem{Hemery2023}
K\'evin H\'emery, Khaldoon Ghanem, Eleanor Crane, Sara~L. Campbell, Joan~M.
  Dreiling, Caroline Figgatt, Cameron Foltz, John~P. Gaebler, Jacob Johansen,
  Michael Mills, Steven~A. Moses, Juan~M. Pino, Anthony Ransford, Mary Rowe,
  Peter Siegfried, Russell~P. Stutz, Henrik Dreyer, Alexander Schuckert, and
  Ramil Nigmatullin.
\newblock ``Measuring the loschmidt amplitude for finite-energy properties of
  the fermi-hubbard model on an ion-trap quantum computer''.
\newblock \href{https://dx.doi.org/10.1103/PRXQuantum.5.030323}{PRX Quantum
  {\bf 5}, 030323}~(2024).

\bibitem{Hofrichter2016}
Christian Hofrichter, Luis Riegger, Francesco Scazza, Moritz H{\"o}fer,
  Diogo~Rio Fernandes, Immanuel Bloch, and Simon F{\"o}lling.
\newblock ``Direct {{Probing}} of the {{Mott Crossover}} in the
  \$\textbackslash mathrm\{\vphantom\}{{SU}}\vphantom\{\}({{N}})\$
  {{Fermi-Hubbard Model}}''.
\newblock \href{https://dx.doi.org/10.1103/PhysRevX.6.021030}{Physical Review X
  {\bf 6}, 021030}~(2016).

\bibitem{Gattringer_IJMFA2016}
Christof Gattringer and Kurt Langfeld.
\newblock ``Approaches to the sign problem in lattice field theory''.
\newblock \href{https://dx.doi.org/10.1142/S0217751X16430077}{International
  Journal of Modern Physics A {\bf 31}, 1643007}~(2016).

\bibitem{Troyer_PRL2005}
Matthias Troyer and Uwe-Jens Wiese.
\newblock ``Computational complexity and fundamental limitations to fermionic
  quantum monte carlo simulations''.
\newblock \href{https://dx.doi.org/10.1103/PhysRevLett.94.170201}{Phys. Rev.
  Lett. {\bf 94}, 170201}~(2005).

\bibitem{QS_roadmap21}
Ehud Altman, Kenneth~R. Brown, Giuseppe Carleo, Lincoln~D. Carr, Eugene Demler,
  Cheng Chin, Brian DeMarco, Sophia~E. Economou, Mark~A. Eriksson, Kai-Mei~C.
  Fu, Markus Greiner, Kaden~R.A. Hazzard, Randall~G. Hulet, Alicia~J. Koll\'ar,
  Benjamin~L. Lev, Mikhail~D. Lukin, Ruichao Ma, Xiao Mi, Shashank Misra,
  Christopher Monroe, Kater Murch, Zaira Nazario, Kang-Kuen Ni, Andrew~C.
  Potter, Pedram Roushan, Mark Saffman, Monika Schleier-Smith, Irfan Siddiqi,
  Raymond Simmonds, Meenakshi Singh, I.B. Spielman, Kristan Temme, David~S.
  Weiss, Jelena Vu\ifmmode \check{c}\else \v{c}\fi{}kovi\ifmmode~\acute{c}\else
  \'{c}\fi{}, Vladan Vuleti\ifmmode~\acute{c}\else \'{c}\fi{}, Jun Ye, and
  Martin Zwierlein.
\newblock ``Quantum simulators: Architectures and opportunities''.
\newblock \href{https://dx.doi.org/10.1103/PRXQuantum.2.017003}{PRX Quantum
  {\bf 2}, 017003}~(2021).

\bibitem{Esslinger_Rev2010}
Tilman Esslinger.
\newblock ``Fermi-hubbard physics with atoms in an optical lattice''.
\newblock
  \href{https://dx.doi.org/10.1146/annurev-conmatphys-070909-104059}{Annual
  Review of Condensed Matter Physics {\bf 1}, 129--152}~(2010).

\bibitem{Bloch_NatPhys2012}
Immanuel Bloch, Jean Dalibard, and Sylvain Nascimbène.
\newblock ``Quantum simulations with ultracold quantum gases''.
\newblock \href{https://dx.doi.org/10.1038/nphys2259}{Nature Physics {\bf 8},
  267--276}~(2012).

\bibitem{Tarruel_Rev2018}
Leticia Tarruell and Laurent Sanchez-Palencia.
\newblock ``Quantum simulation of the hubbard model with ultracold fermions in
  optical lattices''.
\newblock
  \href{https://dx.doi.org/https://doi.org/10.1016/j.crhy.2018.10.013}{Comptes
  Rendus Physique {\bf 19}, 365--393}~(2018).

\bibitem{Borhdt_Rev2021}
Annabelle Bohrdt, Lukas Homeier, Christian Reinmoser, Eugene Demler, and Fabian
  Grusdt.
\newblock ``Exploration of doped quantum magnets with ultracold atoms''.
\newblock
  \href{https://dx.doi.org/https://doi.org/10.1016/j.aop.2021.168651}{Annals of
  Physics {\bf 435}, 168651}~(2021).

\bibitem{Cheuk2015}
Lawrence~W. Cheuk, Matthew~A. Nichols, Melih Okan, Thomas Gersdorf, Vinay~V.
  Ramasesh, Waseem~S. Bakr, Thomas Lompe, and Martin~W. Zwierlein.
\newblock ``Quantum-{{Gas Microscope}} for {{Fermionic Atoms}}''.
\newblock \href{https://dx.doi.org/10.1103/PhysRevLett.114.193001}{Physical
  Review Letters {\bf 114}, 193001}~(2015).

\bibitem{Duarte2015}
Pedro~M. Duarte, Russell~A. Hart, Tsung-Lin Yang, Xinxing Liu, Thereza Paiva,
  Ehsan Khatami, Richard~T. Scalettar, Nandini Trivedi, and Randall~G. Hulet.
\newblock ``Compressibility of a {{Fermionic Mott Insulator}} of {{Ultracold
  Atoms}}''.
\newblock \href{https://dx.doi.org/10.1103/PhysRevLett.114.070403}{Physical
  Review Letters {\bf 114}, 070403}~(2015).

\bibitem{Edge2015}
G.~J.~A. Edge, R.~Anderson, D.~Jervis, D.~C. McKay, R.~Day, S.~Trotzky, and
  J.~H. Thywissen.
\newblock ``Imaging and addressing of individual fermionic atoms in an optical
  lattice''.
\newblock \href{https://dx.doi.org/10.1103/PhysRevA.92.063406}{Physical Review
  A {\bf 92}, 063406}~(2015).

\bibitem{Greif2013}
Daniel Greif, Thomas Uehlinger, Gregor Jotzu, Leticia Tarruell, and Tilman
  Esslinger.
\newblock ``Short-{{Range Quantum Magnetism}} of {{Ultracold Fermions}} in an
  {{Optical Lattice}}''.
\newblock \href{https://dx.doi.org/10.1126/science.1236362}{Science {\bf 340},
  1307--1310}~(2013).

\bibitem{Haller2015}
Elmar Haller, James Hudson, Andrew Kelly, Dylan~A. Cotta, Bruno Peaudecerf,
  Graham~D. Bruce, and Stefan Kuhr.
\newblock ``Single-atom imaging of fermions in a quantum-gas microscope''.
\newblock \href{https://dx.doi.org/10.1038/nphys3403}{Nature Physics {\bf 11},
  738--742}~(2015).

\bibitem{Hart2015}
Russell~A. Hart, Pedro~M. Duarte, Tsung-Lin Yang, Xinxing Liu, Thereza Paiva,
  Ehsan Khatami, Richard~T. Scalettar, Nandini Trivedi, David~A. Huse, and
  Randall~G. Hulet.
\newblock ``Observation of antiferromagnetic correlations in the {{Hubbard}}
  model with ultracold atoms''.
\newblock \href{https://dx.doi.org/10.1038/nature14223}{Nature {\bf 519},
  211--214}~(2015).

\bibitem{Hofstetter2002}
W.~Hofstetter, J.~I. Cirac, P.~Zoller, E.~Demler, and M.~D. Lukin.
\newblock ``High-{{Temperature Superfluidity}} of {{Fermionic Atoms}} in
  {{Optical Lattices}}''.
\newblock \href{https://dx.doi.org/10.1103/PhysRevLett.89.220407}{Physical
  Review Letters {\bf 89}, 220407}~(2002).

\bibitem{Jordens2008}
Robert J{\"o}rdens, Niels Strohmaier, Kenneth G{\"u}nter, Henning Moritz, and
  Tilman Esslinger.
\newblock ``A {{Mott}} insulator of fermionic atoms in an optical lattice''.
\newblock \href{https://dx.doi.org/10.1038/nature07244}{Nature {\bf 455},
  204--207}~(2008).

\bibitem{Messer2015}
Michael Messer, R{\'e}mi Desbuquois, Thomas Uehlinger, Gregor Jotzu, Sebastian
  Huber, Daniel Greif, and Tilman Esslinger.
\newblock ``Exploring {{Competing Density Order}} in the {{Ionic Hubbard
  Model}} with {{Ultracold Fermions}}''.
\newblock \href{https://dx.doi.org/10.1103/PhysRevLett.115.115303}{Physical
  Review Letters {\bf 115}, 115303}~(2015).

\bibitem{Murmann2015}
S.~Murmann, F.~Deuretzbacher, G.~Z{\"u}rn, J.~Bjerlin, S.~M. Reimann,
  L.~Santos, T.~Lompe, and S.~Jochim.
\newblock ``Antiferromagnetic {{Heisenberg Spin Chain}} of a {{Few Cold Atoms}}
  in a {{One-Dimensional Trap}}''.
\newblock \href{https://dx.doi.org/10.1103/PhysRevLett.115.215301}{Physical
  Review Letters {\bf 115}, 215301}~(2015).

\bibitem{Omran2015}
Ahmed Omran, Martin Boll, Timon~A. Hilker, Katharina Kleinlein, Guillaume
  Salomon, Immanuel Bloch, and Christian Gross.
\newblock ``Microscopic {{Observation}} of {{Pauli Blocking}} in {{Degenerate
  Fermionic Lattice Gases}}''.
\newblock \href{https://dx.doi.org/10.1103/PhysRevLett.115.263001}{Physical
  Review Letters {\bf 115}, 263001}~(2015).

\bibitem{Parsons2015}
Maxwell~F. Parsons, Florian Huber, Anton Mazurenko, Christie~S. Chiu, Widagdo
  Setiawan, Katherine {Wooley-Brown}, Sebastian Blatt, and Markus Greiner.
\newblock ``Site-{{Resolved Imaging}} of {{Fermionic}} \$\^\{6\}\textbackslash
  mathrm\{\vphantom\}{{Li}}\vphantom\{\}\$ in an {{Optical Lattice}}''.
\newblock \href{https://dx.doi.org/10.1103/PhysRevLett.114.213002}{Physical
  Review Letters {\bf 114}, 213002}~(2015).

\bibitem{Schneider2008}
U.~Schneider, L.~Hackerm{\"u}ller, S.~Will, {\relax Th}.~Best, I.~Bloch, T.~A.
  Costi, R.~W. Helmes, D.~Rasch, and A.~Rosch.
\newblock ``Metallic and {{Insulating Phases}} of {{Repulsively Interacting
  Fermions}} in a {{3D Optical Lattice}}''.
\newblock \href{https://dx.doi.org/10.1126/science.1165449}{Science {\bf 322},
  1520--1525}~(2008).

\bibitem{Taie2012}
Shintaro Taie, Rekishu Yamazaki, Seiji Sugawa, and Yoshiro Takahashi.
\newblock ``An {{SU}}(6) {{Mott}} insulator of an atomic {{Fermi}} gas realized
  by large-spin {{Pomeranchuk}} cooling''.
\newblock \href{https://dx.doi.org/10.1038/nphys2430}{Nature Physics {\bf 8},
  825--830}~(2012).

\bibitem{Uehlinger2013}
Thomas Uehlinger, Gregor Jotzu, Michael Messer, Daniel Greif, Walter
  Hofstetter, Ulf Bissbort, and Tilman Esslinger.
\newblock ``Artificial {{Graphene}} with {{Tunable Interactions}}''.
\newblock \href{https://dx.doi.org/10.1103/PhysRevLett.111.185307}{Physical
  Review Letters {\bf 111}, 185307}~(2013).

\bibitem{TorstenDanielFermion1}
Torsten~V. Zache, Daniel Gonz{\'{a}}lez-Cuadra, and Peter Zoller.
\newblock ``Fermion-qudit quantum processors for simulating lattice gauge
  theories with matter''.
\newblock \href{https://dx.doi.org/10.22331/q-2023-10-16-1140}{{Quantum} {\bf
  7}, 1140}~(2023).

\bibitem{TorstenDanielFermion2}
Daniel Gonz{\'a}lez-Cuadra, Dolev Bluvstein, Marcin Kalinowski, Raphael
  Kaubruegger, Nishad Maskara, Piero Naldesi, Torsten~V Zache, Adam~M Kaufman,
  Mikhail~D Lukin, Hannes Pichler, et~al.
\newblock ``Fermionic quantum processing with programmable neutral atom
  arrays''.
\newblock
  \href{https://dx.doi.org/https://doi.org/10.1073/pnas.2304294120}{PNAS}~(2023).

\bibitem{arute2020observation}
Frank Arute, Kunal Arya, Ryan Babbush, Dave Bacon, Joseph~C. Bardin, Rami
  Barends, Andreas Bengtsson, Sergio Boixo, Michael Broughton, Bob~B. Buckley,
  David~A. Buell, Brian Burkett, Nicholas Bushnell, Yu~Chen, Zijun Chen, Yu-An
  Chen, Ben Chiaro, Roberto Collins, Stephen~J. Cotton, William Courtney, Sean
  Demura, Alan Derk, Andrew Dunsworth, Daniel Eppens, Thomas Eckl, Catherine
  Erickson, Edward Farhi, Austin Fowler, Brooks Foxen, Craig Gidney, Marissa
  Giustina, Rob Graff, Jonathan~A. Gross, Steve Habegger, Matthew~P. Harrigan,
  Alan Ho, Sabrina Hong, Trent Huang, William Huggins, Lev~B. Ioffe, Sergei~V.
  Isakov, Evan Jeffrey, Zhang Jiang, Cody Jones, Dvir Kafri, Kostyantyn
  Kechedzhi, Julian Kelly, Seon Kim, Paul~V. Klimov, Alexander~N. Korotkov,
  Fedor Kostritsa, David Landhuis, Pavel Laptev, Mike Lindmark, Erik Lucero,
  Michael Marthaler, Orion Martin, John~M. Martinis, Anika Marusczyk, Sam
  McArdle, Jarrod~R. McClean, Trevor McCourt, Matt McEwen, Anthony Megrant,
  Carlos Mejuto-Zaera, Xiao Mi, Masoud Mohseni, Wojciech Mruczkiewicz, Josh
  Mutus, Ofer Naaman, Matthew Neeley, Charles Neill, Hartmut Neven, Michael
  Newman, Murphy~Yuezhen Niu, Thomas~E. O'Brien, Eric Ostby, Bálint Pató,
  Andre Petukhov, Harald Putterman, Chris Quintana, Jan-Michael Reiner, Pedram
  Roushan, Nicholas~C. Rubin, Daniel Sank, Kevin~J. Satzinger, Vadim
  Smelyanskiy, Doug Strain, Kevin~J. Sung, Peter Schmitteckert, Marco Szalay,
  Norm~M. Tubman, Amit Vainsencher, Theodore White, Nicolas Vogt, Z.~Jamie Yao,
  Ping Yeh, Adam Zalcman, and Sebastian Zanker.
\newblock ``Observation of separated dynamics of charge and spin in the
  fermi-hubbard model''~(2020).
\newblock  \href{http://arxiv.org/abs/2010.07965}{arXiv:2010.07965}.

\bibitem{Barends_NatCommun2015}
R.~Barends, L.~Lamata, J.~Kelly, L.~García-Álvarez, A.~G. Fowler, A.~Megrant,
  E.~Jeffrey, T.~C. White, D.~Sank, J.~Y. Mutus, B.~Campbell, Yu~Chen, Z.~Chen,
  B.~Chiaro, A.~Dunsworth, I.-C. Hoi, C.~Neill, P.~J.~J. O’Malley,
  C.~Quintana, P.~Roushan, A.~Vainsencher, J.~Wenner, E.~Solano, and John~M.
  Martinis.
\newblock ``Digital quantum simulation of fermionic models with a
  superconducting circuit''.
\newblock \href{https://dx.doi.org/10.1038/ncomms8654}{Nature Communications
  {\bf 6}, 7654}~(2015).

\bibitem{Sala_PRX2015}
Y.~Salath\'e, M.~Mondal, M.~Oppliger, J.~Heinsoo, P.~Kurpiers,
  A.~Poto\ifmmode~\check{c}\else \v{c}\fi{}nik, A.~Mezzacapo, U.~Las~Heras,
  L.~Lamata, E.~Solano, S.~Filipp, and A.~Wallraff.
\newblock ``Digital quantum simulation of spin models with circuit quantum
  electrodynamics''.
\newblock \href{https://dx.doi.org/10.1103/PhysRevX.5.021027}{Phys. Rev. X {\bf
  5}, 021027}~(2015).

\bibitem{OMalley_PRX2016}
P.~J.~J. O'Malley, R.~Babbush, I.~D. Kivlichan, J.~Romero, J.~R. McClean,
  R.~Barends, J.~Kelly, P.~Roushan, A.~Tranter, N.~Ding, B.~Campbell, Y.~Chen,
  Z.~Chen, B.~Chiaro, A.~Dunsworth, A.~G. Fowler, E.~Jeffrey, E.~Lucero,
  A.~Megrant, J.~Y. Mutus, M.~Neeley, C.~Neill, C.~Quintana, D.~Sank,
  A.~Vainsencher, J.~Wenner, T.~C. White, P.~V. Coveney, P.~J. Love, H.~Neven,
  A.~Aspuru-Guzik, and J.~M. Martinis.
\newblock ``Scalable quantum simulation of molecular energies''.
\newblock \href{https://dx.doi.org/10.1103/PhysRevX.6.031007}{Phys. Rev. X {\bf
  6}, 031007}~(2016).

\bibitem{Stanisic_NatCommun2022}
Stasja Stanisic, Jan~Lukas Bosse, Filippo~Maria Gambetta, Raul~A. Santos,
  Wojciech Mruczkiewicz, Thomas~E. O’Brien, Eric Ostby, and Ashley Montanaro.
\newblock ``Observing ground-state properties of the {Fermi}-{Hubbard} model
  using a scalable algorithm on a quantum computer''.
\newblock \href{https://dx.doi.org/10.1038/s41467-022-33335-4}{Nature
  Communications {\bf 13}, 5743}~(2022).

\bibitem{JW1928}
Pascual Jordan and Eugene~P Wigner.
\newblock ``About the pauli exclusion principle''.
\newblock \href{https://dx.doi.org/https://doi.org/10.1007/BF01331938}{Z. Phys
  {\bf 47}, 14--75}~(1928).

\bibitem{bravyi2022}
Sergey~B. Bravyi and Alexei~Yu. Kitaev.
\newblock ``Fermionic quantum computation''.
\newblock
  \href{https://dx.doi.org/https://doi.org/10.1006/aphy.2002.6254}{Annals of
  Physics {\bf 298}, 210--226}~(2002).

\bibitem{Fradkin80}
Eduardo Fradkin, Mark Srednicki, and Leonard Susskind.
\newblock ``Fermion representation for the ${Z}_{2}$ lattice gauge theory in
  2+1 dimensions''.
\newblock \href{https://dx.doi.org/10.1103/PhysRevD.21.2885}{Phys. Rev. D {\bf
  21}, 2885--2891}~(1980).

\bibitem{Srednicki80}
Mark Srednicki.
\newblock ``Hidden fermions in $z(2)$ theories''.
\newblock \href{https://dx.doi.org/10.1103/PhysRevD.21.2878}{Phys. Rev. D {\bf
  21}, 2878--2884}~(1980).

\bibitem{FermitospinFrank}
F~Verstraete and J~I Cirac.
\newblock ``Mapping local hamiltonians of fermions to local hamiltonians of
  spins''.
\newblock \href{https://dx.doi.org/10.1088/1742-5468/2005/09/p09012}{Journal of
  Statistical Mechanics: Theory and Experiment {\bf 2005},
  P09012--P09012}~(2005).

\bibitem{chen2018}
Yu-An Chen, Anton Kapustin, and {\DJ}or{\dj}e Radi{\v{c}}evi{\'c}.
\newblock ``Exact bosonization in two spatial dimensions and a new class of
  lattice gauge theories''.
\newblock
  \href{https://dx.doi.org/https://doi.org/10.1016/j.aop.2018.03.024}{Annals of
  Physics {\bf 393}, 234--253}~(2018).

\bibitem{kitaev2006}
Alexei Kitaev.
\newblock ``Anyons in an exactly solved model and beyond''.
\newblock
  \href{https://dx.doi.org/https://doi.org/10.1016/j.aop.2005.10.005}{Annals of
  Physics {\bf 321}, 2--111}~(2006).

\bibitem{ErezA}
Erez Zohar and J.~Ignacio Cirac.
\newblock ``Eliminating fermionic matter fields in lattice gauge theories''.
\newblock \href{https://dx.doi.org/10.1103/PhysRevB.98.075119}{Phys. Rev. B
  {\bf 98}, 075119}~(2018).

\bibitem{ErezB}
Erez Zohar and J.~Ignacio Cirac.
\newblock ``Removing staggered fermionic matter in $u(n)$ and $su(n)$ lattice
  gauge theories''.
\newblock \href{https://dx.doi.org/10.1103/PhysRevD.99.114511}{Phys. Rev. D
  {\bf 99}, 114511}~(2019).

\bibitem{ToricCode}
A.Yu. Kitaev.
\newblock ``Fault-tolerant quantum computation by anyons''.
\newblock
  \href{https://dx.doi.org/https://doi.org/10.1016/S0003-4916(02)00018-0}{Annals
  of Physics {\bf 303}, 2--30}~(2003).

\bibitem{chen_PRXQ2023}
Yu-An Chen, Yijia Xu, et~al.
\newblock ``Equivalence between fermion-to-qubit mappings in two spatial
  dimensions''.
\newblock
  \href{https://dx.doi.org/https://doi.org/10.1103/PRXQuantum.4.010326}{PRX
  Quantum {\bf 4}, 010326}~(2023).

\bibitem{PardoPRR2023}
Guy Pardo, Tomer Greenberg, Aryeh Fortinsky, Nadav Katz, and Erez Zohar.
\newblock ``Resource-efficient quantum simulation of lattice gauge theories in
  arbitrary dimensions: Solving for gauss's law and fermion elimination''.
\newblock \href{https://dx.doi.org/10.1103/PhysRevResearch.5.023077}{Phys. Rev.
  Res. {\bf 5}, 023077}~(2023).

\bibitem{MCB_Z2}
Reinis Irmejs, Mari-Carmen Ba\~nuls, and J.~Ignacio Cirac.
\newblock ``Quantum simulation of ${\mathbb{z}}_{2}$ lattice gauge theory with
  minimal resources''.
\newblock \href{https://dx.doi.org/10.1103/PhysRevD.108.074503}{Phys. Rev. D
  {\bf 108}, 074503}~(2023).

\bibitem{Hilker2017}
Timon~A. Hilker, Guillaume Salomon, Fabian Grusdt, Ahmed Omran, Martin Boll,
  Eugene Demler, Immanuel Bloch, and Christian Gross.
\newblock ``Revealing hidden antiferromagnetic correlations in doped hubbard
  chains via string correlators''.
\newblock \href{https://dx.doi.org/10.1126/science.aam8990}{Science {\bf 357},
  484--487}~(2017).

\bibitem{Vijayan_Science2020}
Jayadev Vijayan, Pimonpan Sompet, Guillaume Salomon, Joannis Koepsell, Sarah
  Hirthe, Annabelle Bohrdt, Fabian Grusdt, Immanuel Bloch, and Christian Gross.
\newblock ``Time-resolved observation of spin-charge deconfinement in fermionic
  hubbard chains''.
\newblock
  \href{https://dx.doi.org/https://doi.org/10.1126/science.aay2354}{Science
  {\bf 367}, 186--189}~(2020).

\bibitem{Greiner2021}
Geoffrey Ji, Muqing Xu, Lev~Haldar Kendrick, Christie~S. Chiu, Justus~C.
  Br\"uggenj\"urgen, Daniel Greif, Annabelle Bohrdt, Fabian Grusdt, Eugene
  Demler, Martin Lebrat, and Markus Greiner.
\newblock ``Coupling a mobile hole to an antiferromagnetic spin background:
  Transient dynamics of a magnetic polaron''.
\newblock \href{https://dx.doi.org/10.1103/PhysRevX.11.021022}{Phys. Rev. X
  {\bf 11}, 021022}~(2021).

\bibitem{Bloch2021}
Joannis Koepsell, Dominik Bourgund, Pimonpan Sompet, Sarah Hirthe, Annabelle
  Bohrdt, Yao Wang, Fabian Grusdt, Eugene Demler, Guillaume Salomon, Christian
  Gross, and Immanuel Bloch.
\newblock ``Microscopic evolution of doped mott insulators from polaronic metal
  to fermi liquid''.
\newblock \href{https://dx.doi.org/10.1126/science.abe7165}{Science {\bf 374},
  82--86}~(2021).

\bibitem{Cataldi2024SimulatingSUYangMills}
Giovanni Cataldi, Giuseppe Magnifico, Pietro Silvi, and Simone Montangero.
\newblock ``Simulating (2+1)d su(2) yang-mills lattice gauge theory at finite
  density with tensor networks''.
\newblock \href{https://dx.doi.org/10.1103/PhysRevResearch.6.033057}{Physical
  Review Research {\bf 6}, 033057}~(2024).

\bibitem{LiebBook}
Elliott~H Lieb.
\newblock ``Condensed matter physics and exactly soluble models: selecta of
  elliott h. lieb''.
\newblock
  \href{https://dx.doi.org/https://doi.org/10.1007/978-3-662-06390-3}{Springer
  Science \& Business Media}. ~(2013).

\bibitem{fermionicPEPS}
Philippe Corboz, Rom\'an Or\'us, Bela Bauer, and Guifr\'e Vidal.
\newblock ``Simulation of strongly correlated fermions in two spatial
  dimensions with fermionic projected entangled-pair states''.
\newblock \href{https://dx.doi.org/10.1103/PhysRevB.81.165104}{Phys. Rev. B
  {\bf 81}, 165104}~(2010).

\bibitem{BurrelloZoharPRD}
Erez Zohar and Michele Burrello.
\newblock ``Formulation of lattice gauge theories for quantum simulations''.
\newblock \href{https://dx.doi.org/10.1103/PhysRevD.91.054506}{Phys. Rev. D
  {\bf 91}, 054506}~(2015).

\bibitem{kitaev2009topological}
Alexei Kitaev and Chris Laumann.
\newblock ``Topological phases and quantum computation''.
\newblock
  \href{https://dx.doi.org/https://doi.org/10.48550/arXiv.0904.2771}{Exact
  methods in low-dimensional statistical physics and quantum computingPages
  101--125}~(2009).

\bibitem{Hubbard1963}
John Hubbard.
\newblock ``Electron correlations in narrow energy bands''.
\newblock \href{https://dx.doi.org/10.1098/rspa.1963.0204}{Proceedings of the
  Royal Society of London. Series A. Mathematical and Physical Sciences {\bf
  276}, 238--257}~(1963).

\bibitem{HubbardReview2021}
Daniel~P Arovas, Erez Berg, Steven~A Kivelson, and Srinivas Raghu.
\newblock ``The hubbard model''.
\newblock
  \href{https://dx.doi.org/10.1146/annurev-conmatphys-031620-102024}{Annual
  Review of Condensed Matter Physics{\bf 13}}~(2021).

\bibitem{Lieb_PRL1968}
Elliott~H. Lieb and F.~Y. Wu.
\newblock ``Absence of mott transition in an exact solution of the short-range,
  one-band model in one dimension''.
\newblock \href{https://dx.doi.org/10.1103/PhysRevLett.20.1445}{Phys. Rev.
  Lett. {\bf 20}, 1445--1448}~(1968).

\bibitem{Metzner_PRL1989}
Walter Metzner and Dieter Vollhardt.
\newblock ``Correlated lattice fermions in $d=\ensuremath{\infty}$
  dimensions''.
\newblock \href{https://dx.doi.org/10.1103/PhysRevLett.62.324}{Phys. Rev. Lett.
  {\bf 62}, 324--327}~(1989).

\bibitem{muller-hartmann_correlated_1989}
E.~Müller-Hartmann.
\newblock ``Correlated fermions on a lattice in high dimensions''.
\newblock \href{https://dx.doi.org/10.1007/BF01311397}{Zeitschrift für Physik
  B Condensed Matter {\bf 74}, 507--512}~(1989).

\bibitem{HubbardSim2021}
Mingpu Qin, Thomas Sch{\"a}fer, Sabine Andergassen, Philippe Corboz, and
  Emanuel Gull.
\newblock ``The hubbard model: A computational perspective''.
\newblock
  \href{https://dx.doi.org/10.1146/annurev-conmatphys-090921-033948}{Annual
  Review of Condensed Matter Physics{\bf 13}}~(2021).

\bibitem{LeBlanc_PRX2015}
J.~P.~F. LeBlanc, Andrey~E. Antipov, Federico Becca, Ireneusz~W. Bulik, Garnet
  Kin-Lic Chan, Chia-Min Chung, Youjin Deng, Michel Ferrero, Thomas~M.
  Henderson, Carlos~A. Jim\'enez-Hoyos, E.~Kozik, Xuan-Wen Liu, Andrew~J.
  Millis, N.~V. Prokof'ev, Mingpu Qin, Gustavo~E. Scuseria, Hao Shi, B.~V.
  Svistunov, Luca~F. Tocchio, I.~S. Tupitsyn, Steven~R. White, Shiwei Zhang,
  Bo-Xiao Zheng, Zhenyue Zhu, and Emanuel Gull.
\newblock ``Solutions of the two-dimensional hubbard model: Benchmarks and
  results from a wide range of numerical algorithms''.
\newblock \href{https://dx.doi.org/10.1103/PhysRevX.5.041041}{Phys. Rev. X {\bf
  5}, 041041}~(2015).

\bibitem{LiebHubbard}
Elliott~H. Lieb.
\newblock ``Two theorems on the hubbard model''.
\newblock \href{https://dx.doi.org/10.1103/PhysRevLett.62.1201}{Phys. Rev.
  Lett. {\bf 62}, 1201--1204}~(1989).

\bibitem{LGTN}
Pietro Silvi, Enrique Rico, Tommaso Calarco, and Simone Montangero.
\newblock ``Lattice gauge tensor networks''.
\newblock \href{https://dx.doi.org/10.1088/1367-2630/16/10/103015}{New Journal
  of Physics {\bf 16}, 103015}~(2014).

\bibitem{Cataldi2021a}
Giovanni Cataldi, Ashkan Abedi, Giuseppe Magnifico, Simone Notarnicola,
  Nicola~Dalla Pozza, Vittorio Giovannetti, and Simone Montangero.
\newblock ``Hilbert curve vs {{Hilbert}} space: Exploiting fractal {{2D}}
  covering to increase tensor network efficiency''.
\newblock \href{https://dx.doi.org/10.22331/q-2021-09-29-556}{Quantum {\bf 5},
  556}~(2021).

\bibitem{PhysRevB.105.214201}
Giovanni Ferrari, Giuseppe Magnifico, and Simone Montangero.
\newblock ``Adaptive-weighted tree tensor networks for disordered quantum
  many-body systems''.
\newblock \href{https://dx.doi.org/10.1103/PhysRevB.105.214201}{Phys. Rev. B
  {\bf 105}, 214201}~(2022).

\bibitem{10.21468/SciPostPhysLectNotes.8}
Pietro Silvi, Ferdinand Tschirsich, Matthias Gerster, Johannes Jünemann,
  Daniel Jaschke, Matteo Rizzi, and Simone Montangero.
\newblock ``{The Tensor Networks Anthology: Simulation techniques for many-body
  quantum lattice systems}''.
\newblock \href{https://dx.doi.org/10.21468/SciPostPhysLectNotes.8}{SciPost
  Phys. Lect. NotesPage~8}~(2019).

\bibitem{Montangero_book}
Simone Montangero.
\newblock ``Introduction to tensor network methods''.
\newblock
  \href{https://dx.doi.org/https://doi.org/10.1007/978-3-030-01409-4}{Springer
  International Publishing}. ~(2018).

\bibitem{PhysRevB.93.045116}
Philippe Corboz.
\newblock ``Improved energy extrapolation with infinite projected
  entangled-pair states applied to the two-dimensional hubbard model''.
\newblock \href{https://dx.doi.org/10.1103/PhysRevB.93.045116}{Phys. Rev. B
  {\bf 93}, 045116}~(2016).

\bibitem{nielsen2005}
Michael~A Nielsen et~al.
\newblock ``The fermionic canonical commutation relations and the jordan-wigner
  transform''.
\newblock School of Physical Sciences The University of Queensland {\bf 59},
  75~(2005).
\newblock
  url:~\url{https://futureofmatter.com/assets/fermions_and_jordan_wigner.pdf}.

\bibitem{BRAVYI2002}
Sergey~B. Bravyi and Alexei~Yu. Kitaev.
\newblock ``Fermionic quantum computation''.
\newblock
  \href{https://dx.doi.org/https://doi.org/10.1006/aphy.2002.6254}{Annals of
  Physics {\bf 298}, 210--226}~(2002).

\bibitem{Jiang2020}
Zhang Jiang, Amir Kalev, Wojciech Mruczkiewicz, and Hartmut Neven.
\newblock ``Optimal fermion-to-qubit mapping via ternary trees with
  applications to reduced quantum states learning''.
\newblock \href{https://dx.doi.org/10.22331/q-2020-06-04-276}{{Quantum} {\bf
  4}, 276}~(2020).

\bibitem{Whitfield_PRA2016}
James~D. Whitfield, Vojt\ifmmode \check{e}\else~\v{e}\fi{}ch
  Havl\'{\i}\ifmmode~\check{c}\else \v{c}\fi{}ek, and Matthias Troyer.
\newblock ``Local spin operators for fermion simulations''.
\newblock \href{https://dx.doi.org/10.1103/PhysRevA.94.030301}{Phys. Rev. A
  {\bf 94}, 030301}~(2016).

\bibitem{Steudtner_PRA2019}
Mark Steudtner and Stephanie Wehner.
\newblock ``Quantum codes for quantum simulation of fermions on a square
  lattice of qubits''.
\newblock \href{https://dx.doi.org/10.1103/PhysRevA.99.022308}{Phys. Rev. A
  {\bf 99}, 022308}~(2019).

\bibitem{Setia_PRR2019}
Kanav Setia, Sergey Bravyi, Antonio Mezzacapo, and James~D. Whitfield.
\newblock ``Superfast encodings for fermionic quantum simulation''.
\newblock \href{https://dx.doi.org/10.1103/PhysRevResearch.1.033033}{Phys. Rev.
  Res. {\bf 1}, 033033}~(2019).

\bibitem{Jiang_PRApp2019}
Zhang Jiang, Jarrod McClean, Ryan Babbush, and Hartmut Neven.
\newblock ``Majorana loop stabilizer codes for error mitigation in fermionic
  quantum simulations''.
\newblock \href{https://dx.doi.org/10.1103/PhysRevApplied.12.064041}{Phys. Rev.
  Appl. {\bf 12}, 064041}~(2019).

\bibitem{Derby_PRB2021}
Charles Derby, Joel Klassen, Johannes Bausch, and Toby Cubitt.
\newblock ``Compact fermion to qubit mappings''.
\newblock \href{https://dx.doi.org/10.1103/PhysRevB.104.035118}{Phys. Rev. B
  {\bf 104}, 035118}~(2021).

\bibitem{Kollath_PRL2005}
C.~Kollath, U.~Schollw\"ock, and W.~Zwerger.
\newblock ``Spin-charge separation in cold fermi gases: A real time analysis''.
\newblock \href{https://dx.doi.org/10.1103/PhysRevLett.95.176401}{Phys. Rev.
  Lett. {\bf 95}, 176401}~(2005).

\bibitem{Grusdt2018}
F.~Grusdt, M.~K\'anasz-Nagy, A.~Bohrdt, C.~S. Chiu, G.~Ji, M.~Greiner,
  D.~Greif, and E.~Demler.
\newblock ``Parton theory of magnetic polarons: Mesonic resonances and
  signatures in dynamics''.
\newblock \href{https://dx.doi.org/10.1103/PhysRevX.8.011046}{Phys. Rev. X {\bf
  8}, 011046}~(2018).

\bibitem{qmatchatea_0_5_2}
Marco Ballarin, Daniel Jaschke, and Montangero Simone.
\newblock ``Quantum tea: qmatchatea''.
\newblock \href{https://dx.doi.org/10.5281/zenodo.11619266}{Zenodo}~(2024).

\bibitem{gottesman1997}
Daniel Gottesman.
\newblock ``Stabilizer codes and quantum error correction''.
\newblock
  \href{https://dx.doi.org/https://doi.org/10.48550/arXiv.quant-ph/9705052}{California
  Institute of Technology}. ~(1997).

\bibitem{Rudolph2024}
Manuel~S Rudolph, Jing Chen, Jacob Miller, Atithi Acharya, and Alejandro
  Perdomo-Ortiz.
\newblock ``Decomposition of matrix product states into shallow quantum
  circuits''.
\newblock \href{https://dx.doi.org/10.1088/2058-9565/ad04e6}{Quantum Science
  and Technology {\bf 9}, 015012}~(2023).

\bibitem{wauters2020}
Matteo~M. Wauters, Glen~B. Mbeng, and Giuseppe~E. Santoro.
\newblock ``Polynomial scaling of the quantum approximate optimization
  algorithm for ground-state preparation of the fully connected $p$-spin
  ferromagnet in a transverse field''.
\newblock \href{https://dx.doi.org/10.1103/PhysRevA.102.062404}{Phys. Rev. A
  {\bf 102}, 062404}~(2020).

\bibitem{bluvstein2024}
Dolev Bluvstein, Simon~J Evered, Alexandra~A Geim, Sophie~H Li, Hengyun Zhou,
  Tom Manovitz, Sepehr Ebadi, Madelyn Cain, Marcin Kalinowski, Dominik
  Hangleiter, et~al.
\newblock ``Logical quantum processor based on reconfigurable atom arrays''.
\newblock
  \href{https://dx.doi.org/https://doi.org/10.1038/s41586-023-06927-3}{Nature
  {\bf 626}, 58--65}~(2024).

\bibitem{dasilva2024}
MP~da~Silva, C~Ryan-Anderson, JM~Bello-Rivas, A~Chernoguzov, JM~Dreiling,
  C~Foltz, JP~Gaebler, TM~Gatterman, D~Hayes, N~Hewitt, et~al.
\newblock ``Demonstration of logical qubits and repeated error correction with
  better-than-physical error rates''~(2024).
\newblock  \href{http://arxiv.org/abs/2404.02280}{arXiv:2404.02280}.

\bibitem{ATTN}
Timo Felser, Simone Notarnicola, and Simone Montangero.
\newblock ``Efficient tensor network ansatz for high-dimensional quantum
  many-body problems''.
\newblock \href{https://dx.doi.org/10.1103/PhysRevLett.126.170603}{Phys. Rev.
  Lett. {\bf 126}, 170603}~(2021).

\bibitem{iPEPSalgo}
J.~Jordan, R.~Or\'us, G.~Vidal, F.~Verstraete, and J.~I. Cirac.
\newblock ``Classical simulation of infinite-size quantum lattice systems in
  two spatial dimensions''.
\newblock \href{https://dx.doi.org/10.1103/PhysRevLett.101.250602}{Phys. Rev.
  Lett. {\bf 101}, 250602}~(2008).

\bibitem{code}
Marco Ballarin, Giovanni Cataldi, Giuseppe Magnifico, Daniel Jaschke, Marco
  Di~Liberto, Ilaria Siloi, Simone Montangero, and Pietro Silvi.
\newblock ``Simulation scripts for "digital quantum simulation of lattice
  fermion theories with local encoding"''.
\newblock \href{https://dx.doi.org/10.5281/zenodo.11619804}{Zenodo}~(2024).

\bibitem{qtealeaves_0_5_12}
Marco Ballarin, Giovanni Cataldi, Aurora Costantini, Daniel Jaschke, Giuseppe
  Magnifico, Simone Montangero, Simone Notarnicola, Alice Pagano, Luka Pavesic,
  Marco Rigobello, Nora Reinić, Simone Scarlatella, and Pietro Silvi.
\newblock ``Quantum tea: qtealeaves''.
\newblock \href{https://dx.doi.org/10.5281/zenodo.10498929}{Zenodo}~(2024).

\bibitem{jaschke2022}
Daniel Jaschke and Simone Montangero.
\newblock ``Is quantum computing green? an estimate for an energy-efficiency
  quantum advantage''.
\newblock \href{https://dx.doi.org/10.1088/2058-9565/acae3e}{Quantum Science
  and Technology {\bf 8}, 025001}~(2023).

\end{thebibliography}

\end{document}